\documentclass[a4paper]{article}
\usepackage[T1]{fontenc}
\usepackage{authblk}
\usepackage{graphicx}

\title{Characterization of microbulk detectors in argon- and neon-based mixtures}
\author[1]{F.J.~Iguaz}
\author[1]{E.~Ferrer-Ribas}
\author[1]{A.~Giganon}
\author[1]{I.~Giomataris}
\affil[1]{IRFU, Centre d'\'Etudes Nucl\'eaires de Saclay (CEA-Saclay), Gif-sur-Yvette, France}
\begin{document}
\maketitle

\begin{abstract}
A recent Micromegas manufacturing technique, so called Microbulk, has been developed, improving the uniformity and stability of this kind of detectors. Excellent energy resolutions have been obtained, reaching values as low as 11\% FWHM at 5.9 keV in Ar+5\%iC$_4$H$_{10}$. This detector has other advantages like its flexible structure, low material budget and high radio-purity. Two microbulk detectors with gaps of 50 and 25~$\mu$m have been characterized in argon- and neon-based mixtures with ethane, isobutane and cyclohexane. The results will be presented and discussed. The gain curves have been fitted to the Rose-Korff gain model and dependences of the electron mean free path and the threshold energy for ionization have been obtained. The possible relation between these two parameters and the energy resolution will be also discussed.
\end{abstract}

\section{Introduction}
Micromegas (for MICRO MEsh GAseous Structure) is a parallel-plate detector invented by I. Giomataris in 1995 \cite{Giomataris:1995fq}. It consists of a thin metallic grid (commonly called mesh) and an anode plane, separated by insullated pillars. Both structures define a small gap (between 20 and 300~$\mu$m), where primary electrons generated in the conversion volume are amplified, applying moderate voltages at the cathode and the mesh. This technology has proven to have many advantages like its high granularity, good energy and time resolution, easy construction, little mass and gain stability.

\medskip
The first detectors were built by screwing two different frames: the anode plane and the metallic grid, under which the insullated pillars were electroformed. By applying voltages to both structures, the intense electric field pulled down the mesh and the flatness was thus defined by the height of the pillars, with an accuracy better than 10~$\mu$m. The good flatness and parallelism between the anode and the mesh was obtained only if the delicate operation of screwing was successful. To avoid this operation, two different technologies (called bulk \cite{Giomataris:2006yg} and microbulk \cite{Adriamonje:2010sa, Iguaz:2011fi}) have been developped so that the readout plane and the mesh formed a single integrated structure.

\medskip
In the microbulk technology, the raw material is a thin flexible polyimide foil with a thin copper layer on each side. The foil is glued on top of a rigid substrate that provides the support of the micro-structure and carries the anode strips or pixels. In the first step, a thin photoresistive film is laminated on top of the kapton foil and is insolated by UV light to produce the required mask. The copper is then removed by a standard lithographic process, generating the pattern of a thin mesh. The polyimide between the mesh and the readout plane is then etched and removed under the mesh holes, but it remains almost intact elsewhere \cite{Adriamonje:2010sa}.

\medskip
The resulting detector has a more homogeneous amplification gap and a better mesh than the classical Micromegas, reducing the avalanche fluctuations and improving the energy resolution. There is also a reduction of the border regions, eliminating dead zones and allowing its application in imaging with x-rays or neutrons. Moreover, the detector is made of kapton and copper, two of the materials with the best levels of radiopurity. These materials define a level of contamination below PMTs\cite{Cebrian:2011sc}. The only possible drawbacks of this technique are its lack of robustness and the limited maximum size ($30 \times 30$ cm$^2$), due to the actual equipment. Nevertheless, the improved detectors have been applied in several experiments like nTOF \cite{Colonna:2011nc}, CAST \cite{SAune:2009sa} or NEXT\cite{Cebrian:2010sc2, Dafni:2010td}.

\medskip
The microbulk technology shows the best energy resolution of MPGD detectors, as good as 11\%~FWHM at 5.9 keV in Ar+5\%iC$_4$H$_{10}$, improving the values around 14\%~FWHM reached by proportional counters two decades ago \cite{Agrawal:1988pca}. The reference gas in tests was fixed by studies of that time, as a high gain and a good energy resolution was obtained. However, the improved performances of microbulk detectors give the oportunity to study in detail the influence of the gas mixture on these parameters and to verify the conclusions obtained with proportional counters. Moreover, if microbulk detectors could work at higher gains ($> 10^4$) keeping a good energy resolution, this technology could be used in sub-keV applications, like in Synchroton Radiation and Dark Matter searches where the low energy threshold is crutial.

\medskip
In this article, we present the performance of microbulk detectors in argon and neon-based mixtures. The detectors tested respectively have a gap size of 50 and 25 $\mu$m. The first gap is the more common used in the new microbulk detectors and the second one is expected to show better performances at high pressure gases \cite{Giomataris:1998yg}. In the second section, the setup and the procedure are described in detail. The features of microbulk detectors in argon- and neon-based mixtures with three different quenchers are described in sections three and four. Finally, we present the Rose-Korff gain model for Micromegas, the dependencies obtained for some gas parameters and we discuss the validity of this model and its possible relation with the energy resolution. We finish this report with some conclusions and an outlook of the actual developments.

\section{Setup and procedure description}
\begin{figure}[htb!]
\centering
\includegraphics[width=70mm]{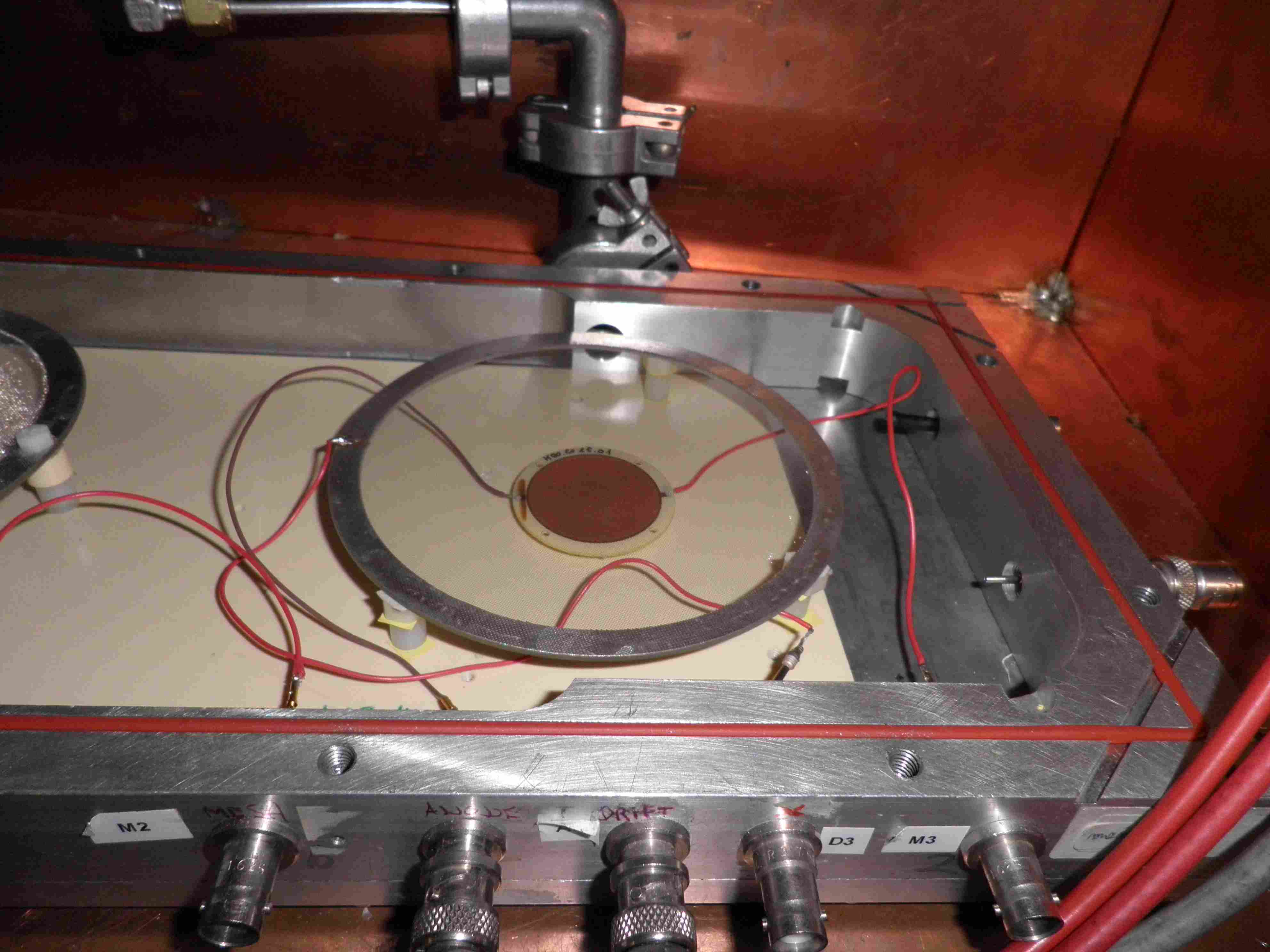}
\hspace{5mm}
\includegraphics[width=40mm]{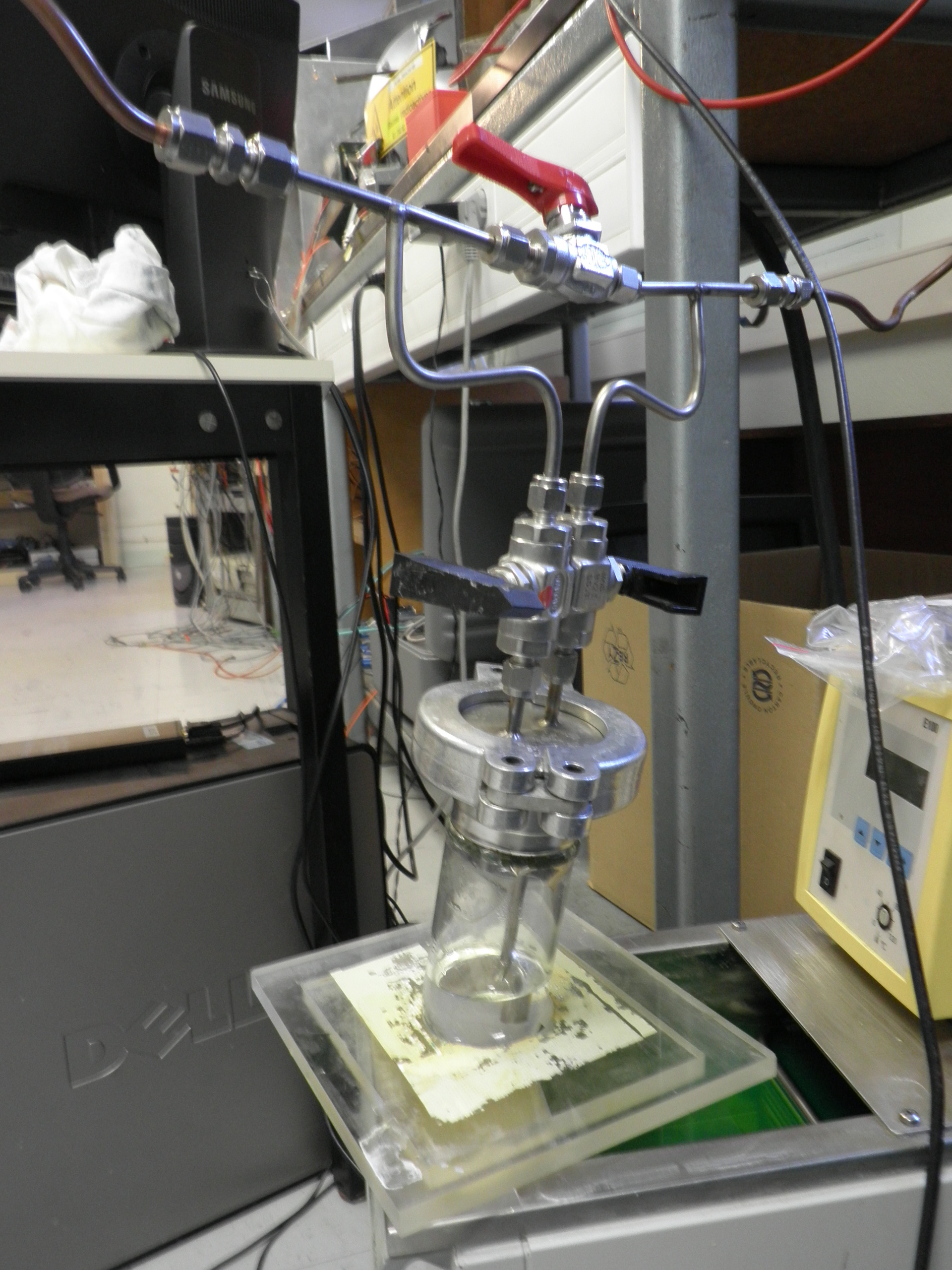}
\caption{\it Left: A view of one microbulk detector installed in the vessel. Right: A view of the gas setup used for the cyclohexane mixtures. The base gas is forced to pass by a glass vessel filled with the liquid quencher. The container is situated inside a refrigerator, shown opened under the vessel.}
\label{fig:setup2}
\end{figure}

The vessel used in these tests (figure \ref{fig:setup2}, left) was specifically designed to characterize a maximum of three micromegas detectors (not necessary microbulk) in the same gas conditions, with a maximum surface of $50 \times 50$ mm$^2$. It consists of an aluminium box, with an internal volume of $60 \times 165 \times 30$ mm$^3$ and a wall thickness of 8 mm. The chamber is equipped with two gas entrances to circulate the desired gas. Detectors are placed on a plastic support situated inside the vessel, which electrically isolates them from the inner walls. For each detector, there is a mesh frame screwed to the plastic plate, which is used as a drift cathode and defines a drift distance of 10~mm. The drift, mesh and anode voltages are respectively extracted by a high voltage (SHV) and two signal (BNC) feed-throughs for each detector. On the top cap of the chamber, several holes allow the calibration by x-ray sources situated outside. These holes are covered by an aluminized mylar films, that are scotched to the inner wall of the cap with kapton and create the gas-tightness of the vessel.

\medskip
During at least one hour, a flow of the desired mixture was circulated by the chamber. In the case of mixtures with cyclohexane, the base gas was forced to pass by a glass vessel, filled with this liquid quencher (figure \ref{fig:setup2}, right). The gas concentration was defined by the liquid temperature, which was kept constant by a refrigerator in which the vessel was immersed. This temperature could not be higher than the ambient one to avoid possible quencher condensations inside the vessel. For a temperature of 25$^\mathrm{o}$C, this condition fixes a maximum concentration of 11\%.

\medskip
The detectors were illuminated by an iron $^{55}$Fe source (x-rays of 5.9 keV) keeping the same gas flow. The mesh voltage was typically varied from 200 to 600~V and the drift one from 280 to 5000~V, depending on the type of gas. Both voltages were powered independently by a CAEN N471A module. The avalanche induces a negative signal in the mesh, which is read out by an ORTEC 142C preamplifier. The preamplifier output was fed into an ORTEC 472A spectroscopy amplifier and subsequently into a multichannel analyzer AMPTEK MCA-8000A for spectra building. For reducing systematic errors, each spectrum has a minimum of $5 \times 10^4$ points.

\medskip
Two microbulk detectors respectively with a gap size of 25 and 50~$\mu$m have been characterized. They have a circular shape (35 mm of diameter) and a single non-segmented anode. Their specific references are M50.70.35.04 and M25.50.25.09, which respectively correspond to a pitch distance of 70 and 50~$\mu$m and a hole diameter of 35 and 25~$\mu$m. These particular detectors have showed an energy resolution of 11.7\% and 11.8\%~FWHM at 5.9 keV in Ar~+~5\%~iC$_4$H$_{10}$, values which are close to the best value observed for microbulk detectors (11\% FWHM).

\section{Characterization in argon-based mixtures}
\begin{figure}[htb!]
\centering
\includegraphics[width=75mm]{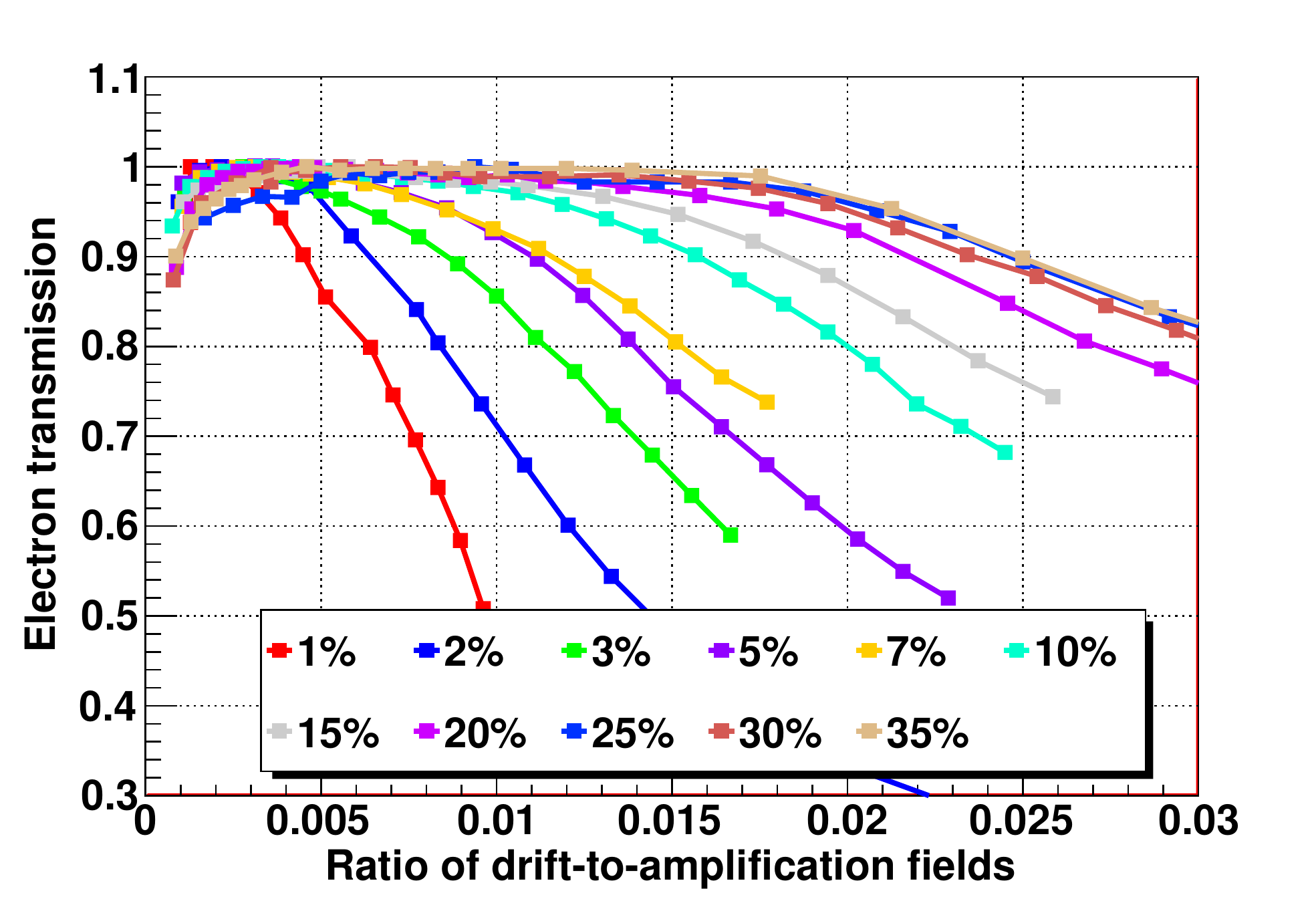}
\includegraphics[width=75mm]{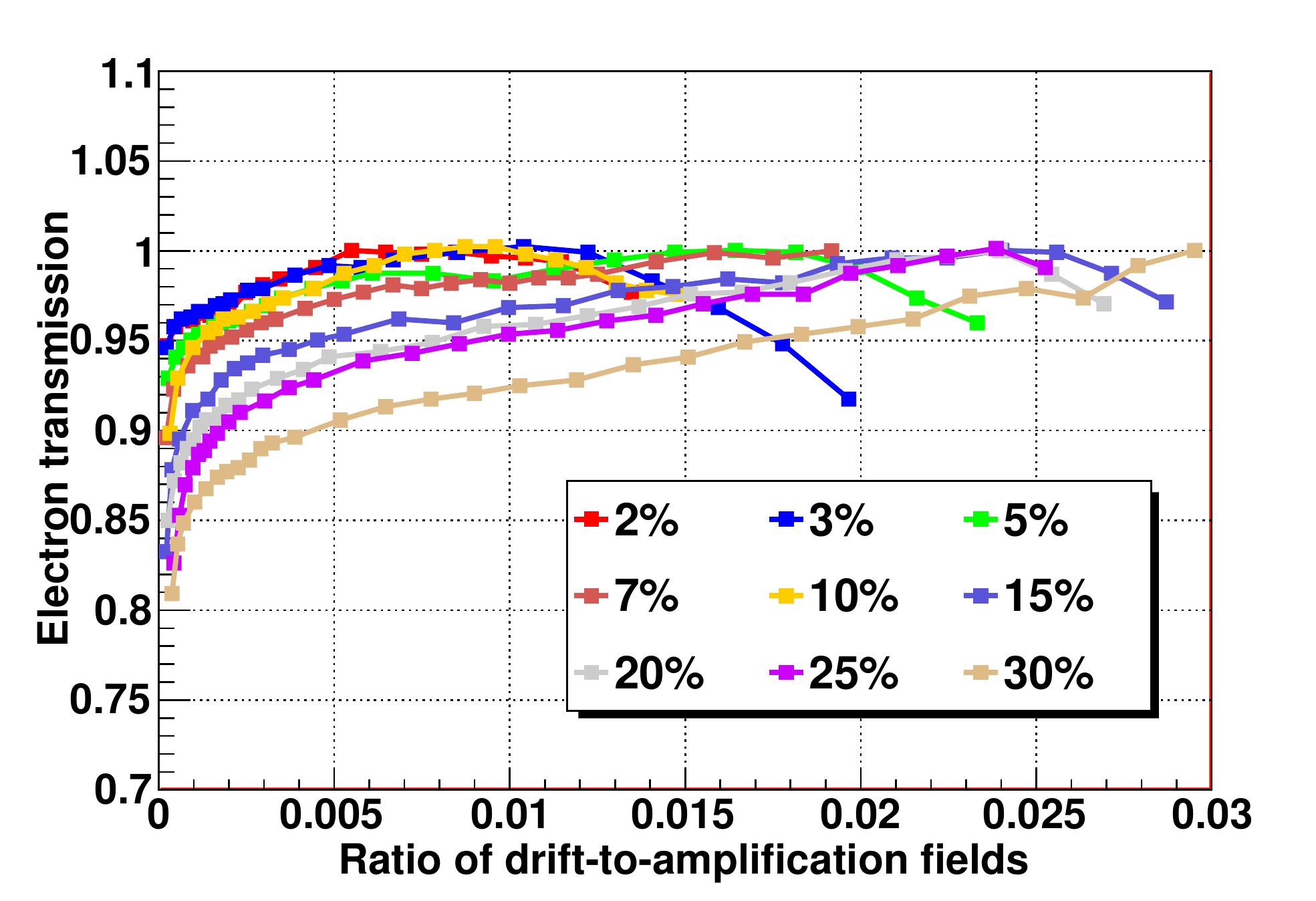}
\caption{\it Dependence of the electron transmission with the ratio of drift-to-amplification fields for two microbulk detectors with gaps of 50 and 25 $\mu$m in argon-isobutane mixtures. The gain has been normalized to the maximum of each series. The percentage of each series corresponds to the isobutane concentration.}
\label{fig:TransArIso}
\end{figure}

The drift voltage was firstly varied in our tests for a fixed mesh voltage to obtain the electron transmission curve, that shows the mesh transparency to primary ionization. These curves are shown for the argon-isobutane case and both detectors in figure \ref{fig:TransArIso}. Microbulk detectors show a plateau of maximum electron transmission for a wide range of ratios of drift-to-amplification fields. At higher drift fields, the mesh stops being transparent for the primary electrons generated in the conversion volume and both the gain and the energy resolution degrade. The specific maximum value for the ratio depends on the diffusion coefficient, being wider the plateau for higher quantities of the quencher. This fact is in agreement with the shorter plateaus observed in pure gases \cite{Dafni:2009df, Iguaz:2010fj} and the wider plateaus measured in gases with low diffusion coefficients \cite{Chefdeville:2009mc}.

\begin{figure}[htb!]
\centering
\includegraphics[width=75mm]{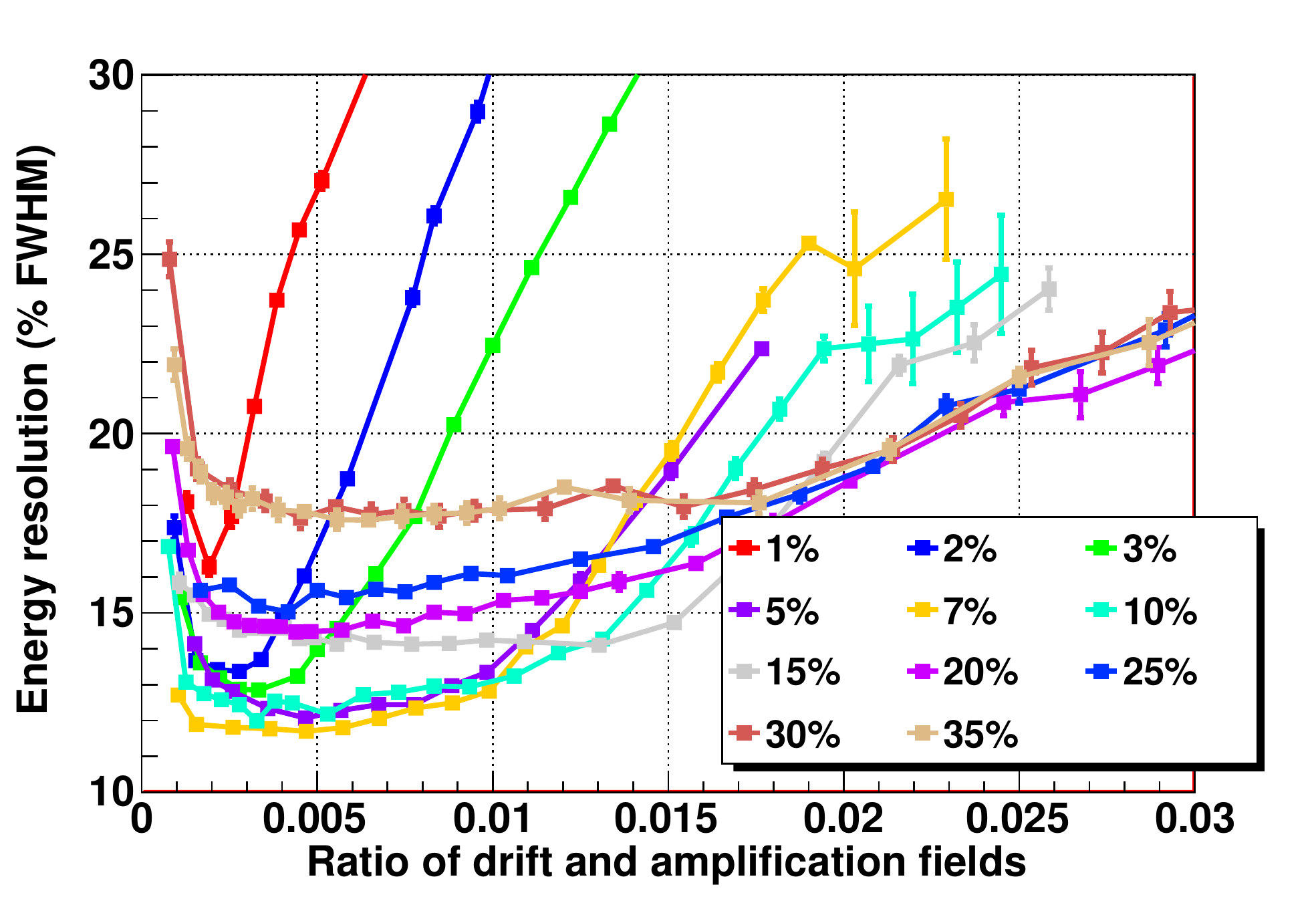}
\includegraphics[width=75mm]{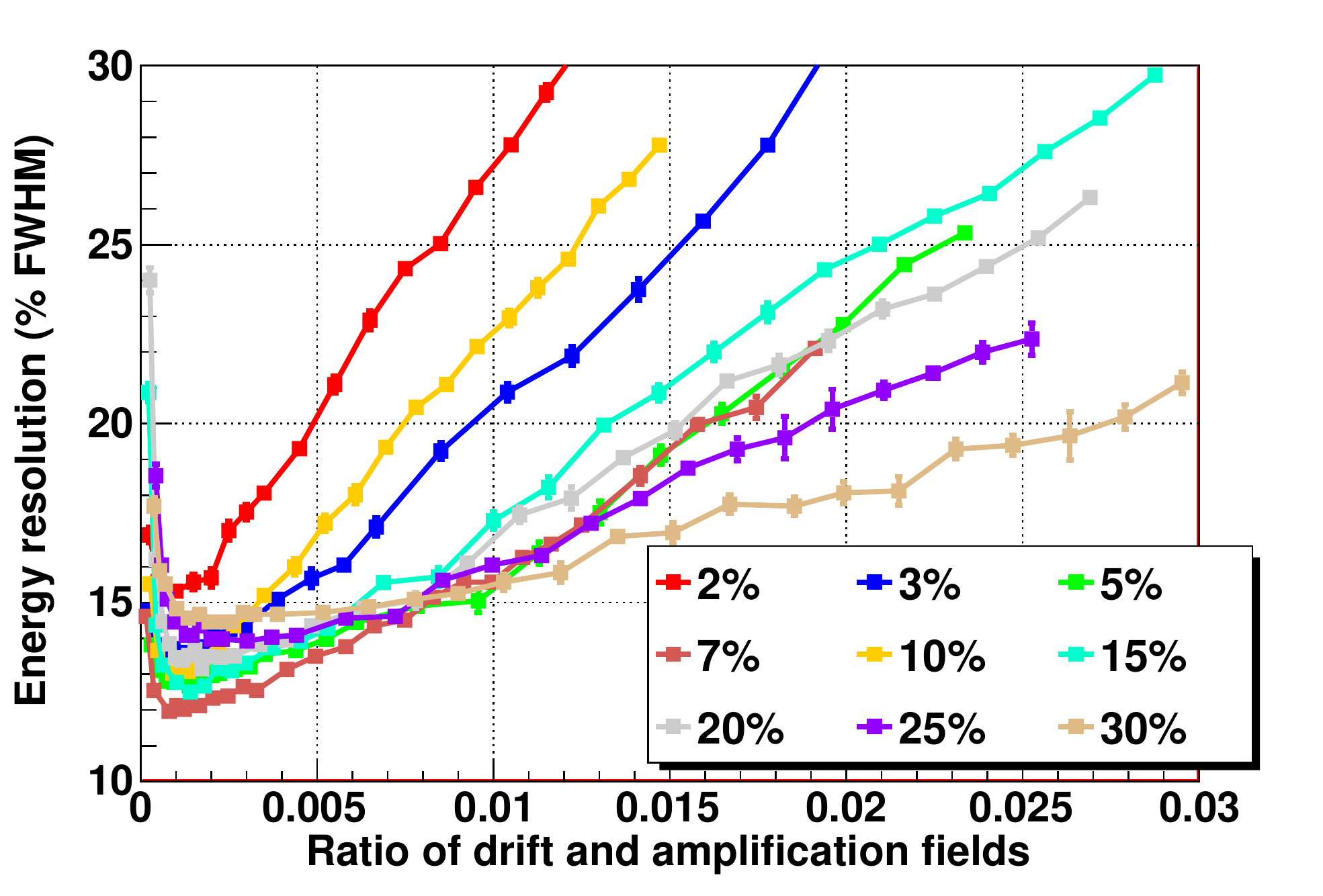}
\caption{\it Dependence of the energy resolution with the ratio of drift-to-amplification fields for two microbulk detectors with gaps of 50 and 25 $\mu$m in argon-isobutane mixtures. The percentage of each series corresponds to the isobutane concentration.}
\label{fig:EResDriftArIso}
\end{figure}

\medskip
However, the detector of 25~$\mu$m-thickness gap does not show the same behaviour, as shown in figure \ref{fig:TransArIso} (right). For low isobutane concentrations, there is a short plateau of maximum mesh transparency but it starts at relative high drift field. At high isobutane concentrations, this plateau disappears and there is a steady increase of the detector gain with the drift field. This effect has been verified by other microbulk of the same gap, a hole pitch between 50 and 100 $\mu$m and a hole diameter between 20 and 30 $\mu$m. A possible explanation is that the mesh may not be completely transparent for electrons for the pitch and hole diameter considered. Indeed, the best values of the energy resolution are reached at low drift fields (figure \ref{fig:EResDriftArIso}, right).

\medskip
The dependence of the energy resolution with the ratio of fields is shown in figure \ref{fig:EResDriftArIso} for argon-isobutane mixtures. For a gap size of 50~$\mu$m, the energy resolution shows its best value for the same drift fields for which the electron transmission is the optimum. There is also a dependence on the isobutane concentration. For values below 5\%, the energy resolution is worse as not all the UV photons produced in the avalanche are absorbed \cite{Bronic:1998ikb}. For values between 5 and 7\%, the energy resolution reaches its optimum and then degrades at higher isobutane concentrations. There is an increase of the gain fluctuations due to a higher proportion of scattering processes than ionization ones \cite{Schindler:2010hs}.

\medskip
In the case of the 25 $\mu$m-thickness-gap detector, the energy resolution reaches its best values for the lowest drift fields. This fact supports the hypothesis that the mesh is not totally transparent for electrons. There is also a dependence on the isobutane concentrations, but not so strong as for the former detector. The energy resolution degrades from 12\% up to 14.5\%~FWHM when the isobutane concentration passes from 7\% to 30\%, which is less than the 19\% FWHM measured with the 50~$\mu$m-thickness-gap detector.

\begin{figure}[htb!]
\centering
\includegraphics[width=75mm]{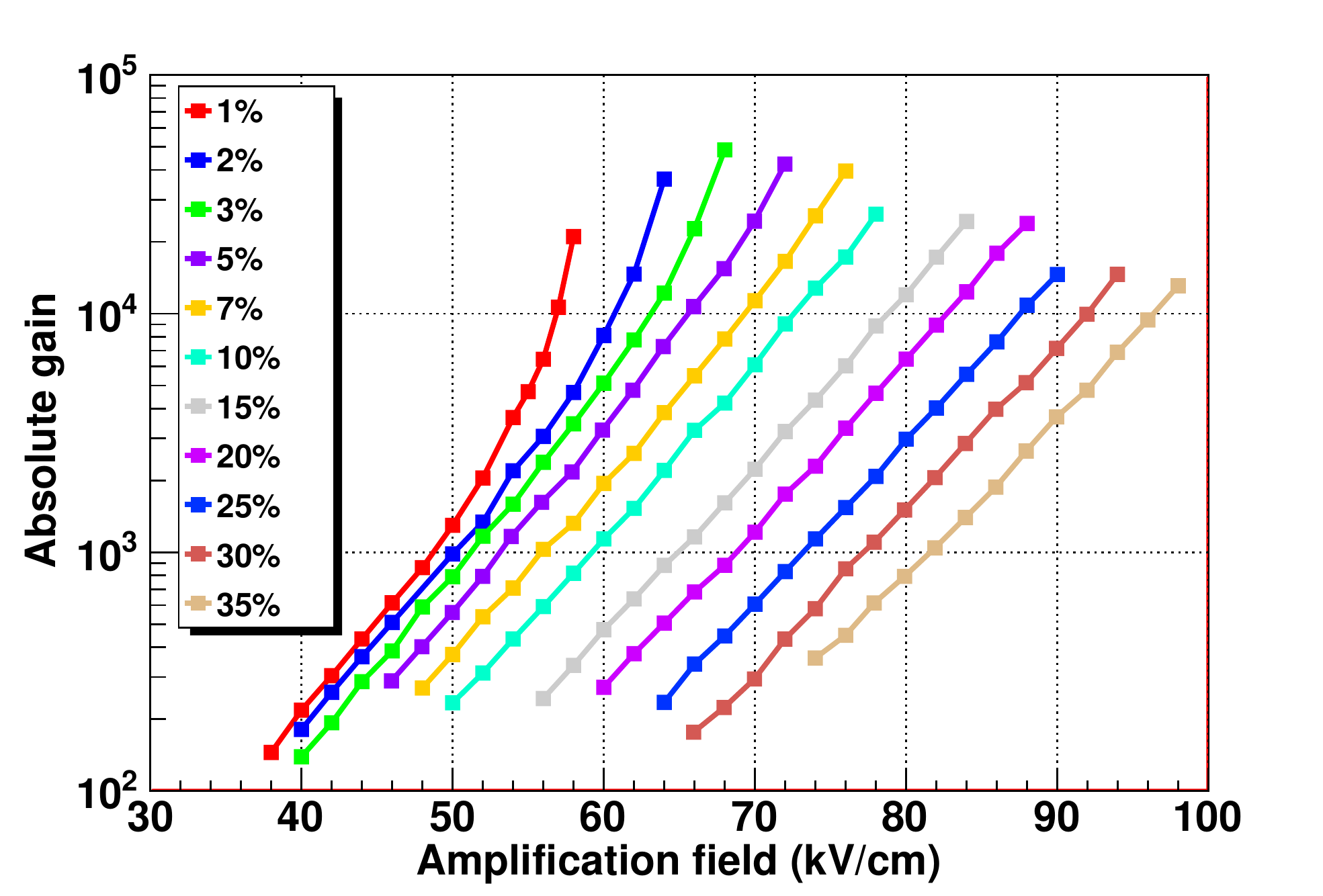}
\includegraphics[width=75mm]{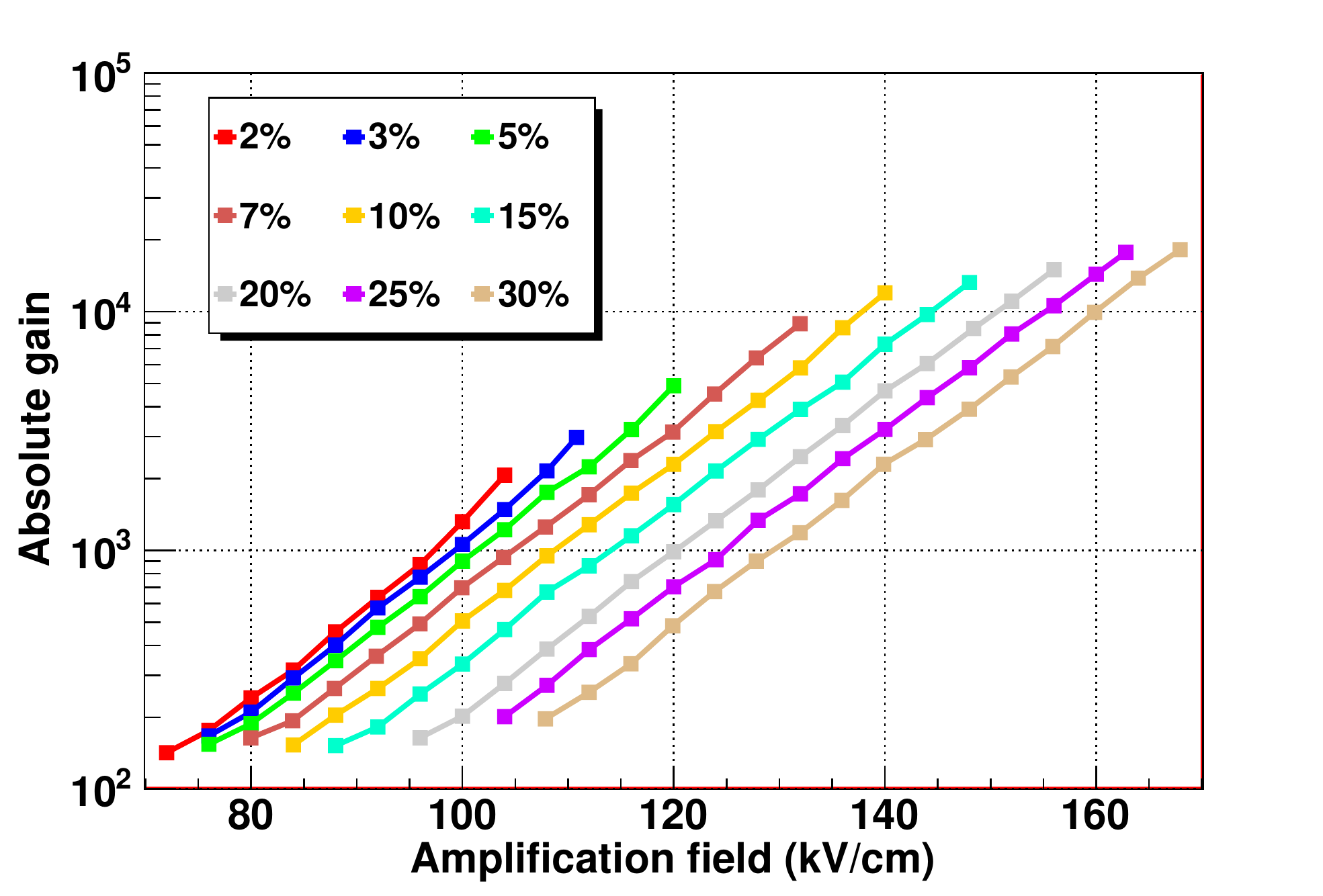}
\caption{\it Dependence of the absolute gain with the amplification field for two microbulk detectors with gaps of 50 (left) and 25 $\mu$m (right) in argon-isobutane mixtures. The maximum gain of each curve was obtained just before the spark limit. The percentage of each series corresponds to the isobutane concentration.}
\label{fig:GainArIso}
\end{figure}

\medskip
The dependence of the peak position with the mesh voltage generates the gain curves, shown in figure \ref{fig:GainArIso} for argon-isobutane mixtures. The maximum gain before the spark limit is respectively $2 \times 10^4$ and $10^4$ for the detectors with a gap size of 50 and 25$~\mu$m. This fact happens for all mixtures in the first case and only for isobutane concentrations over 10\% for the second one. Apart from that, the gain curves for low quantities of isobutane show a deviation from the Rose and Korff gain model \cite{Rose:1941mr} at high amplification fields. This over-exponential behaviour is due to the low quencher concentrations, which cannot avoid the production of secondary avalanches by the UV photons generated in the primary avalanche \cite{Bronic:1998ikb}.

\begin{figure}[htb!]
\centering
\includegraphics[width=75mm]{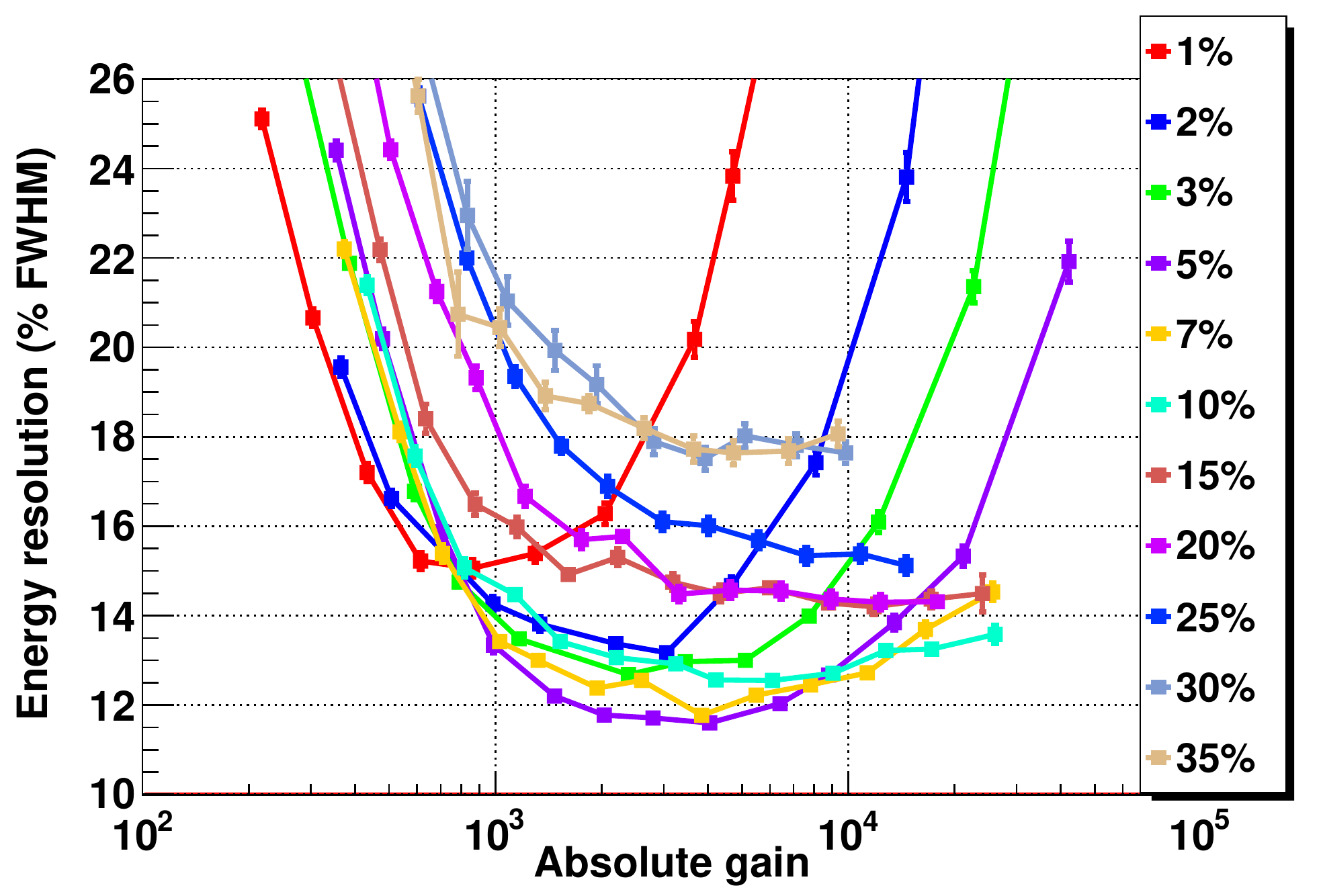}
\includegraphics[width=75mm]{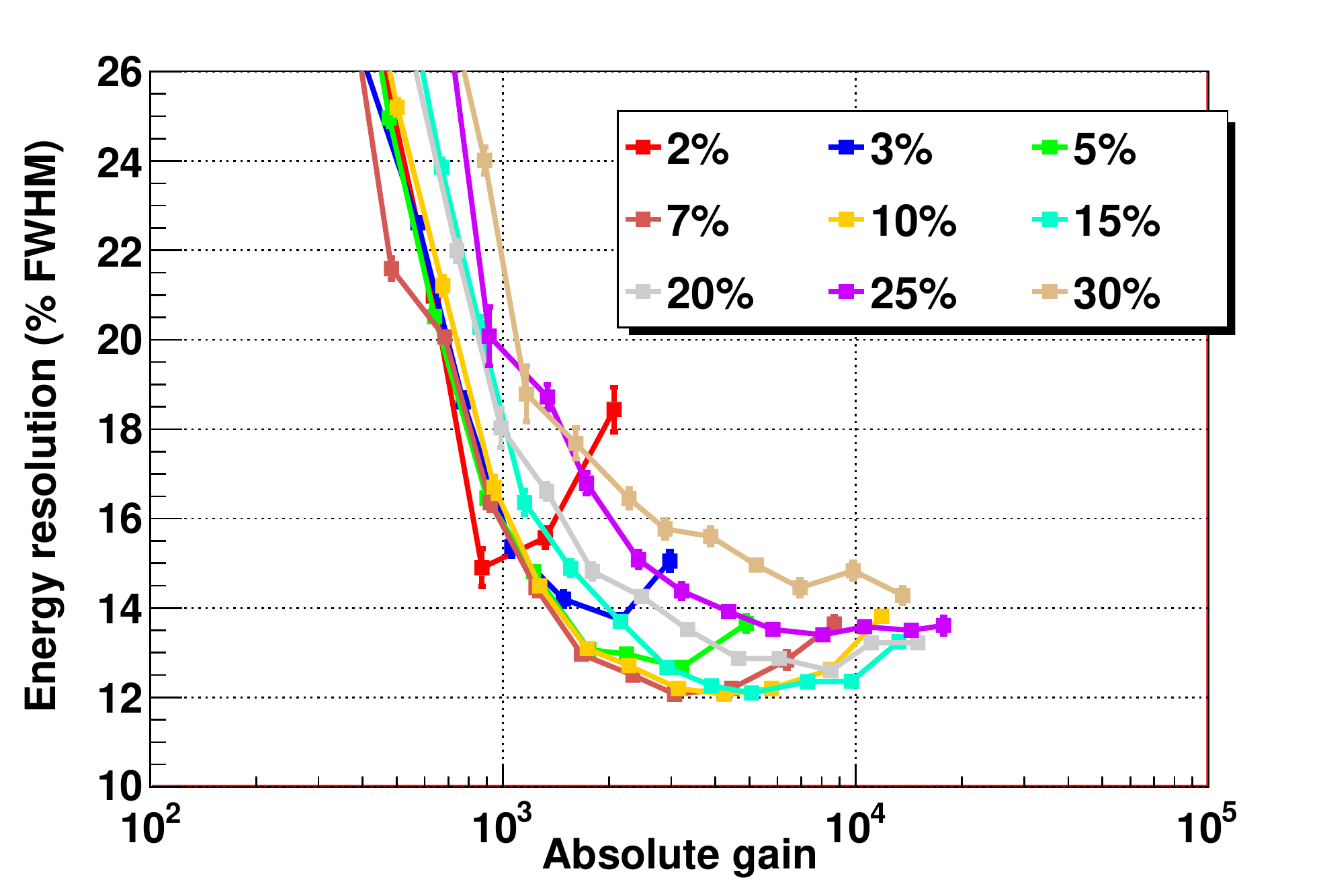}
\caption{\it Dependence of the energy resolution with the absolute gain for two detectors of 50 (left) and 25~$\mu$m-thickness-gap (right) in argon-isobutane mixtures. The maximum gain of each curve was obtained just before the spark limit. The percentage of each series corresponds to the isobutane concentration.}
\label{fig:EResGainArIso}
\end{figure}

\medskip
The energy resolution has also a dependence with the amplification field (figure \ref{fig:EResGainArIso}), and more specifically with the gain, which was already observed with proportional counters \cite{Agrawal:1988pca, Agrawal:1989pca}. At low gains, the energy resolution degrades because the signal starts being comparable to the noise. At high gains, the resolution worsens due to the increase of the gain fluctuations. This effect disappears at high isobutane concentrations. However, the energy resolution is also worse, due to the larger gain fluctuations \cite{Schindler:2010hs}.

\begin{figure}[htb!]
\centering
\includegraphics[width=75mm]{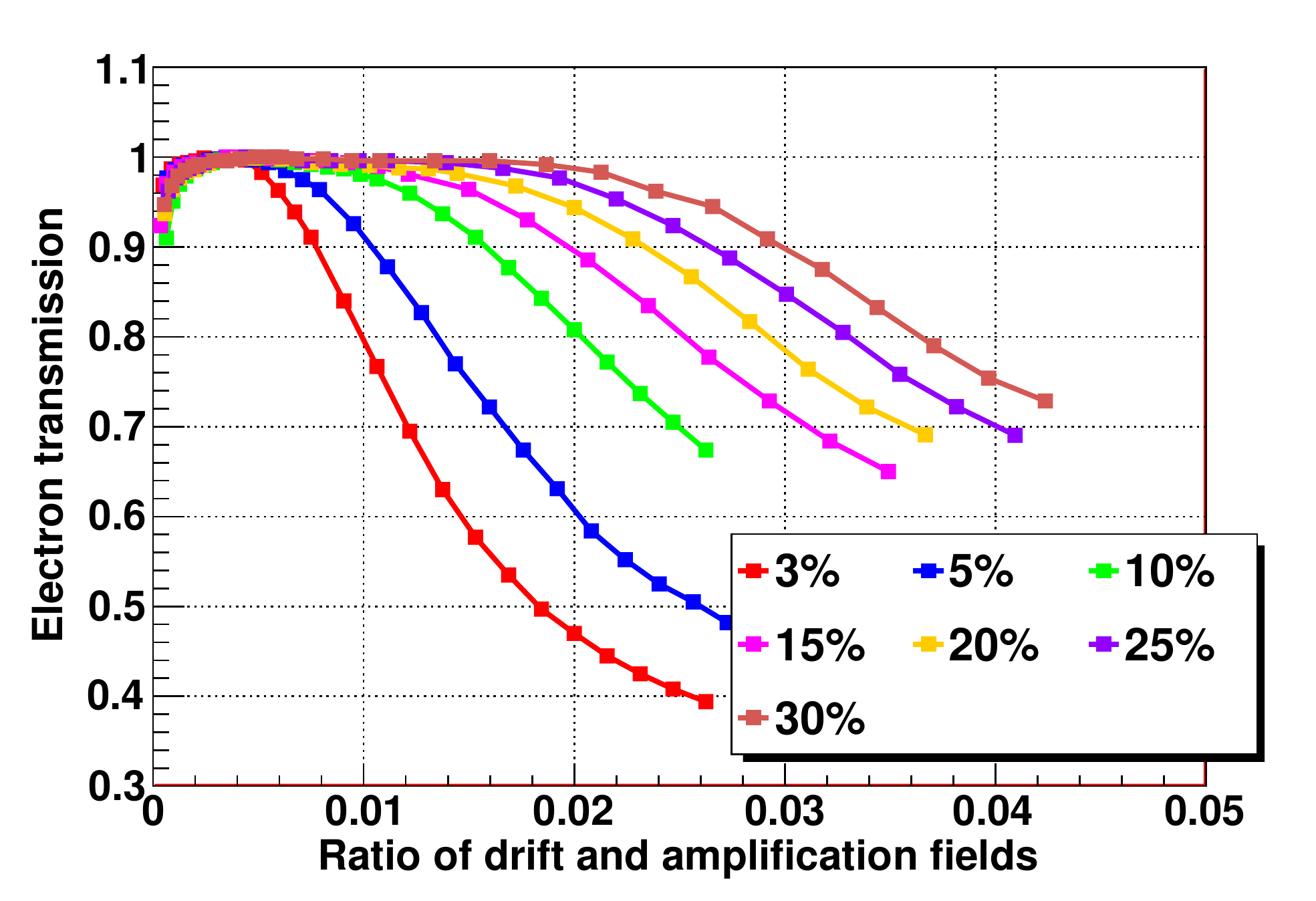}
\includegraphics[width=75mm]{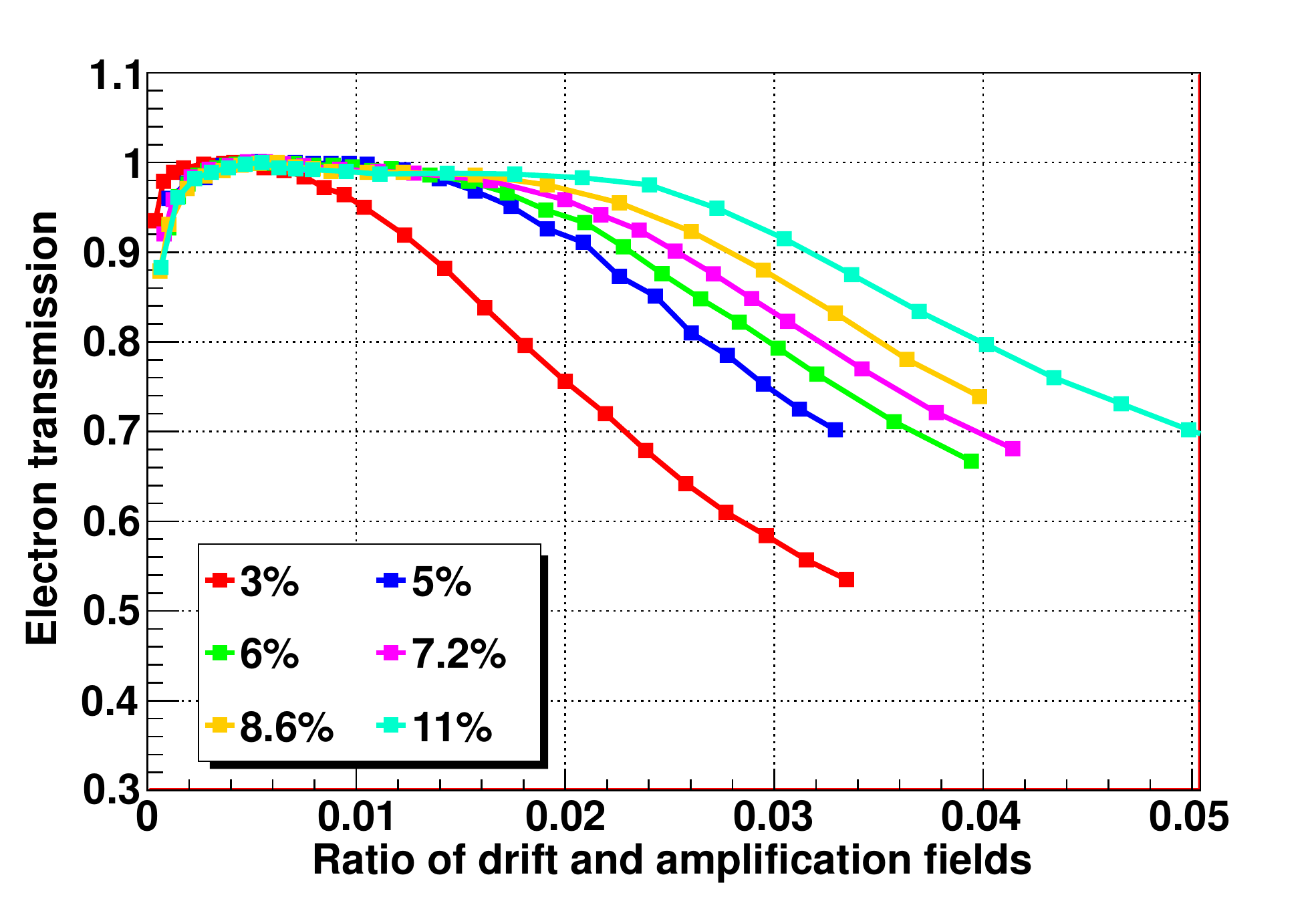}
\caption{\it Dependence of the electron transmission with the ratio of drift and amplification fields for a detector of 50~$\mu$m-thickness-gap in argon-ethane (left) and -cyclohexane mixtures (right). The gain has been normalized to the maximum of each series. The percentage of each series corresponds to the quencher concentration.}
\label{fig:TransM50Argon}
\end{figure}

\medskip
The results obtained for other quenchers are similar to the ones presented for argon-isobutane. We will just focus on the main differences. The first particularity (shown in figure \ref{fig:TransM50Argon}) is that the plateau of maximum electron transmission is wider for mixtures with cyclohexane and narrower for ethane, in comparison to isobutane. The argon-ethane result is consistent with its higher diffusion coefficient for the same percentage of quencher \cite{Magboltz}. In the case of cyclohexane, no data is available on its gas properties.

\begin{figure}[htb!]
\centering
\includegraphics[width=75mm]{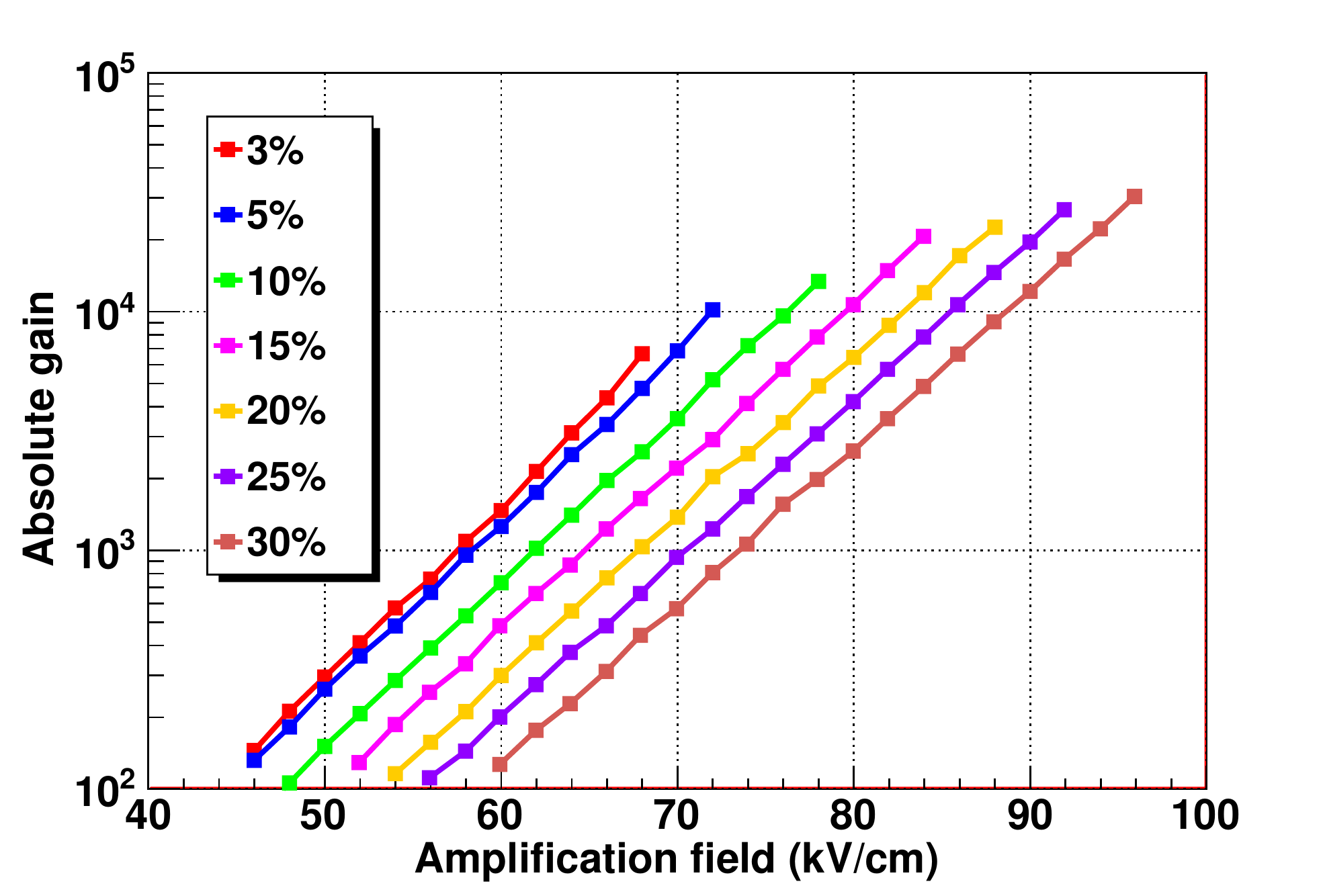}
\includegraphics[width=75mm]{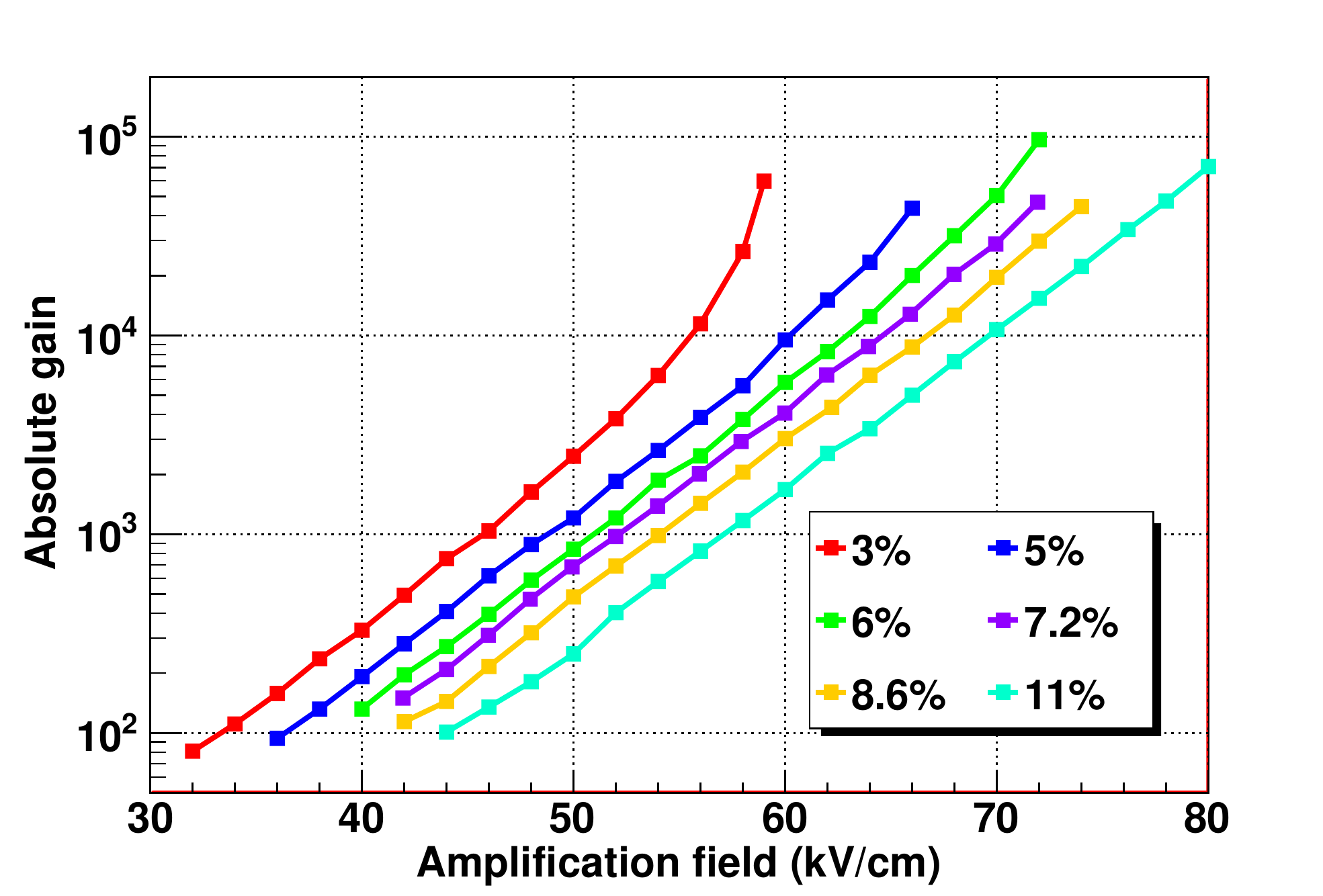}
\caption{\it Dependence of the absolute gain with the absolute gain for a detector of 50 $\mu$m-thickness-gap in mixtures of argon with ethane (left) and cyclohexane (right). The maximum gain of each curve was obtained just before the spark limit. The percentage of each series corresponds to the quencher concentration.}
\label{fig:GainArCycloEth}
\end{figure}

\medskip
As shown in figure \ref{fig:GainArCycloEth}, the maximum gain before the spark limit is respectively $2 \times 10^4$ and $4 \times 10^4$ for mixtures with ethane and cyclohexane and the 50 $\mu$m-thickness-gap detector. Only for cyclohexane, gains higher than in argon-isobutane mixtures are reached. We have also observed that the required voltage to reach the same gain is lower for cyclohexane, than for isobutane or ethane. For instance, an amplification field of 65~kV/cm must be applied to reach a gain of $10^4$ in Ar+5\%iC$_4$H$_10$. This field is only 61~kV/cm in the case of cyclohexane and increases up to 72~kV/cm for ethane.

\medskip
In table \ref{tab:eresargon} we show the best values for the energy resolution obtained in each kind of mixture. We first note that the optimum performance of the 50~$\mu$m-thickness-gap microbulk is reached with quencher concentrations around 5\%. For the gap of 25~$\mu$m, this percentage rises up to 10\%. Nevertheless, the results are quite similar for all mixtures. From some measurements made with proportional counters in \cite{Agrawal:1988pca}, it was concluded that quenchers whose ionization threshold were similar to the first metastable level of argon (11.5 eV) should show a higher gain and a better energy resolution. This improvement is related to a higher Penning effect \cite{Bronic:1996ikb}, i.e., an energy transfer from the base gas atom to the quencher molecule, resulting in an increase of the avalanche ionization. However, this fact does not agree in the case of cyclohexane, as even if its ionization threshold (9.9 eV) is a bit low, the gain is higher with no degradation on the energy resolution.

\medskip
Recent studies \cite{Sahin:2010os} support the main role of the non-metastable levels in Penning effect. Quenchers whose ionization threshold is lower than this level should show the same energy resolution. The higher gain observed in argon-cyclohexane mixtures could be related to a little contribution of argon excimers, created for energies between 8.9 and 10.3 eV.

\begin{table}[htb!]
\begin{center}
$$
\begin{array}{cc|cc|cc}
\multicolumn{1}{c}{}&\multicolumn{1}{c}{}&\multicolumn{2}{c}{\mbox{50~$\mu$m gap}}&\multicolumn{2}{c}{\mbox{25~$\mu$m gap}}\\
{\mbox{Type of}}&{\mbox{Ion. thres}}&{\mbox{Resolution}}&{\mbox{Quencher}}&{\mbox{Resolution}}&{\mbox{Quencher}}\\
{\mbox{quencher}}&{\mbox{eV}}&{\mbox{\% FWHM}}&{\%}&{\mbox{\% FWHM}}&{\%}\\
\hline
{\mbox{Ethane}}&{11.7}&{12.0}&{5-10}&{11.9}&{10-25}\\
{\mbox{Isobutane}}&{10.8}&{11.7}&{5-7}&{12.1}&{7-15}\\
{\mbox{Cyclohexane}}&{9.9}&{12.0}&{5-6}&{11.6}&{7.2-11}\\
\end{array}
$$
\end{center}
\caption{\it Best values of the energy resolution in mixtures of argon with ethane, isobutane and cyclohexane for two microbulk detectors with gaps of 50 and 25~$\mu$m. The range of quencher concentration for which these values are obtained is also shown. The systematic error of all values is 0.2\% FWHM.}
\label{tab:eresargon}
\end{table}

\begin{figure}[htb!]
\centering
\includegraphics[width=75mm]{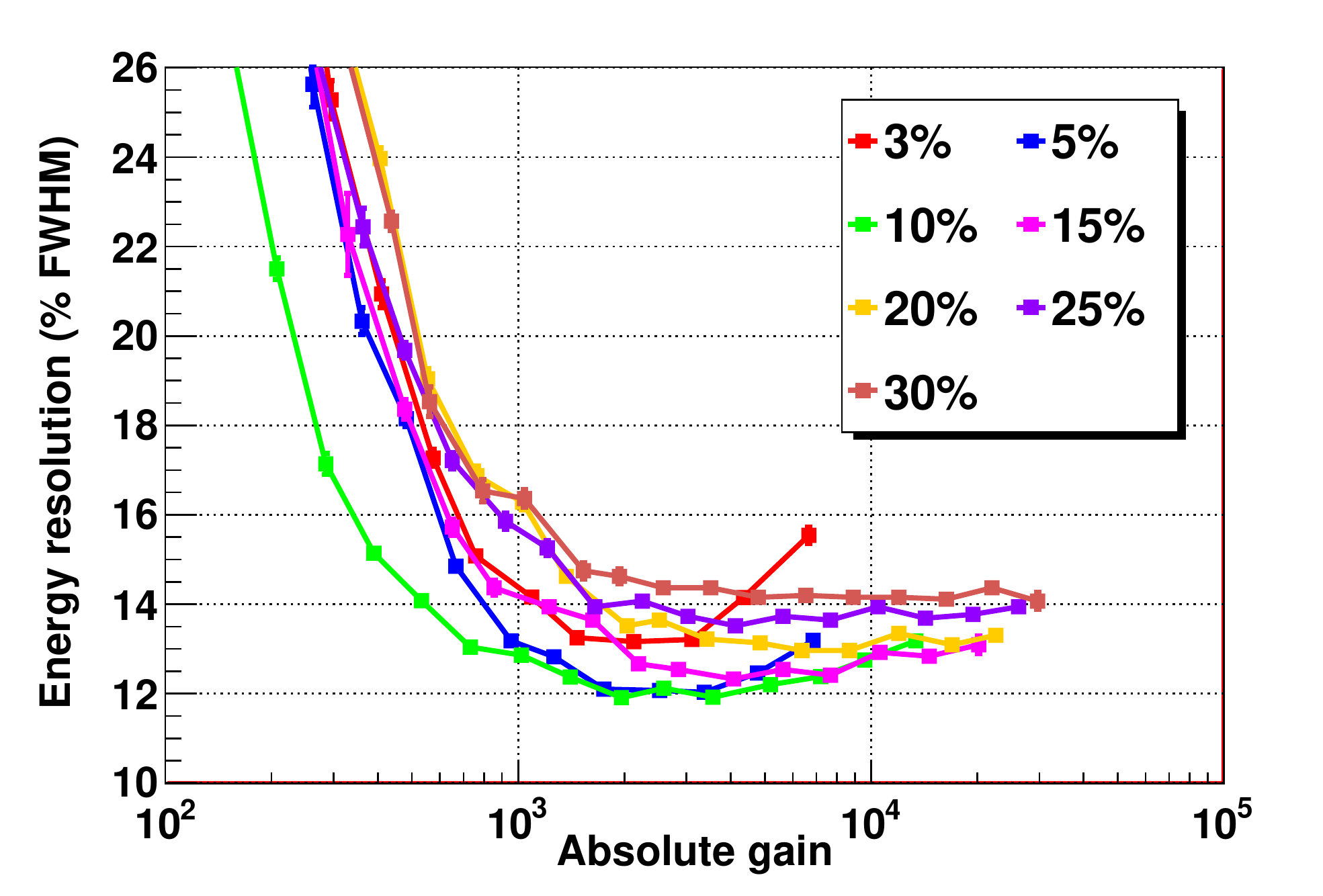}
\includegraphics[width=75mm]{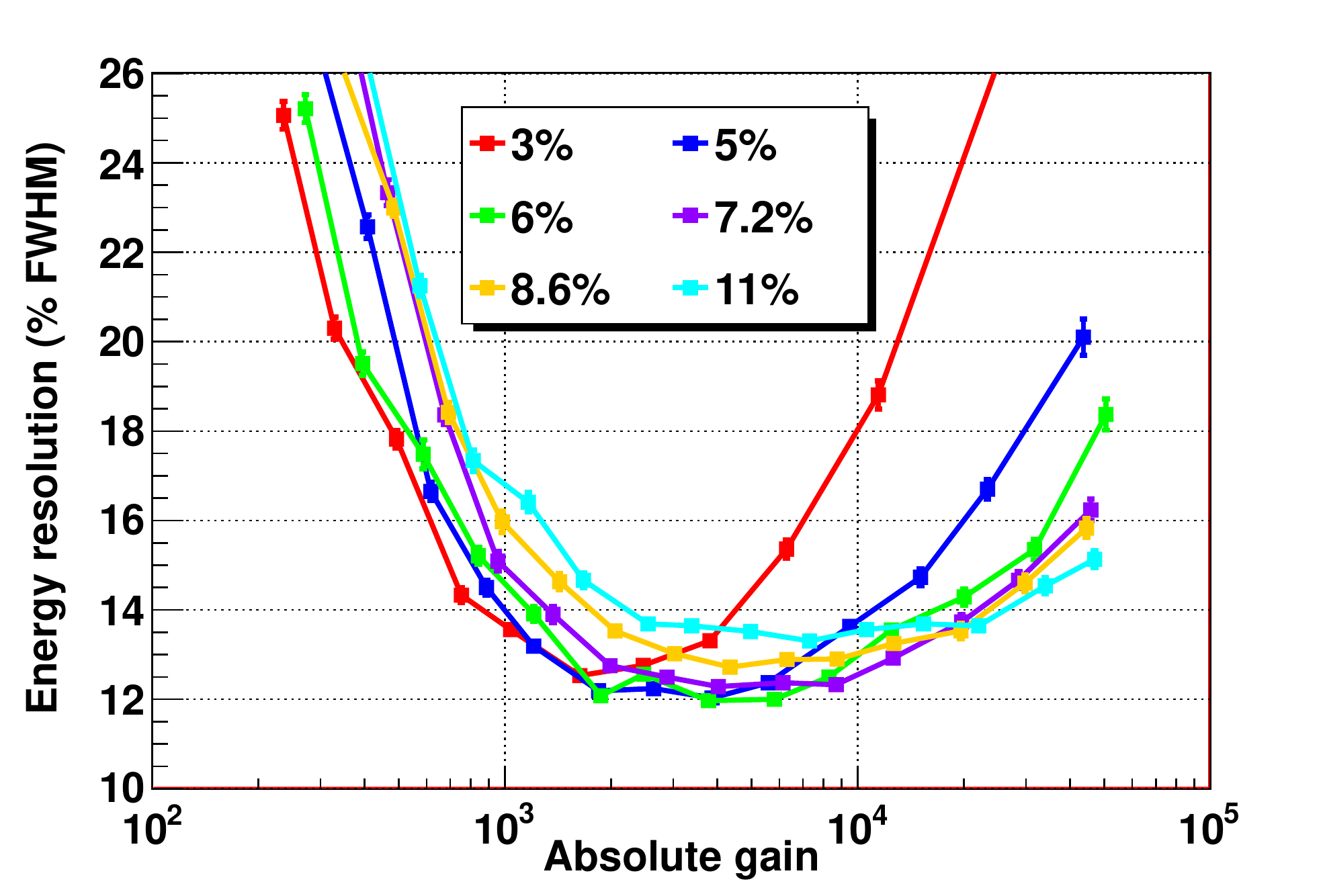}
\caption{\it Dependence of the energy resolution with the absolute gain for a detector of 50~$\mu$m-thickness-gap in argon-ethane (left) and -cyclohexane mixtures (right). The maximum gain of each curve was obtained just before the spark limit. The percentage of each series corresponds to the quencher concentration.}
\label{fig:EResGainM50}
\end{figure}

\medskip
We finally present in figure \ref{fig:EResGainM50} the dependence of the energy resolution with the gain for the microbulk of 50~$\mu$m-thickness-gap in mixtures of argon with ethane and cyclohexane. The maximum gain before the spark limit is respectively $2 \times 10^4$ and $4 \times 10^4$. Note that in the case of cyclohexane there is a clear degradation of the energy resolution at high gains, which is absent for ethane.

\section{Neon-based mixtures for sub-keV applications}
Micromegas detectors have been typically operated in argon-isobutane mixtures between 2 and 5\% for a wide range of applications. This gas is well adapted for measurements in the 1-10 keV range, providing an excellent energy resolution and gains up to $2 \times 10^4$. Other gas mixtures are being studied to increase micromegas sensitivity in the sub-keV region. For this purpose, the signal to noise ratio must be increased and higher gains are needed, keeping a good energy resolution. A possible solution is the replacement of the base gas by neon, as the total charge per single avalanche is increased and approaches the Rather limit ($\approx 10^8$ electrons). However, the efficiency of the photon absorption is lower in this kind of mixtures, passing from 70\% down to 10\% at 6 keV, which limits the range of applications. We present the main features of the two detectors studied in neon mixtures with the same quenchers used before (isobutane, cyclohexane and ethane).

\begin{figure}[htb!]
\centering
\includegraphics[width=75mm]{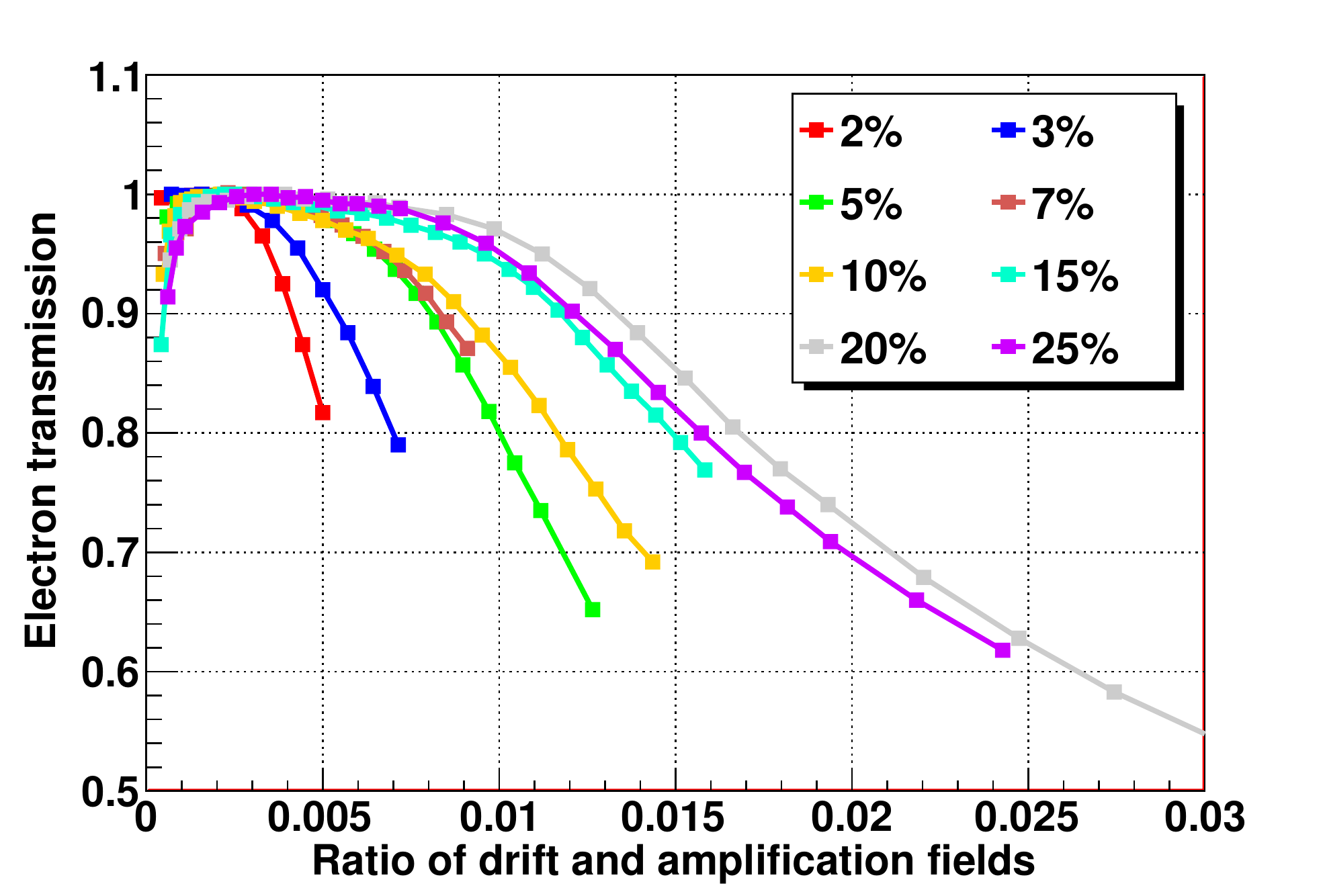}
\includegraphics[width=75mm]{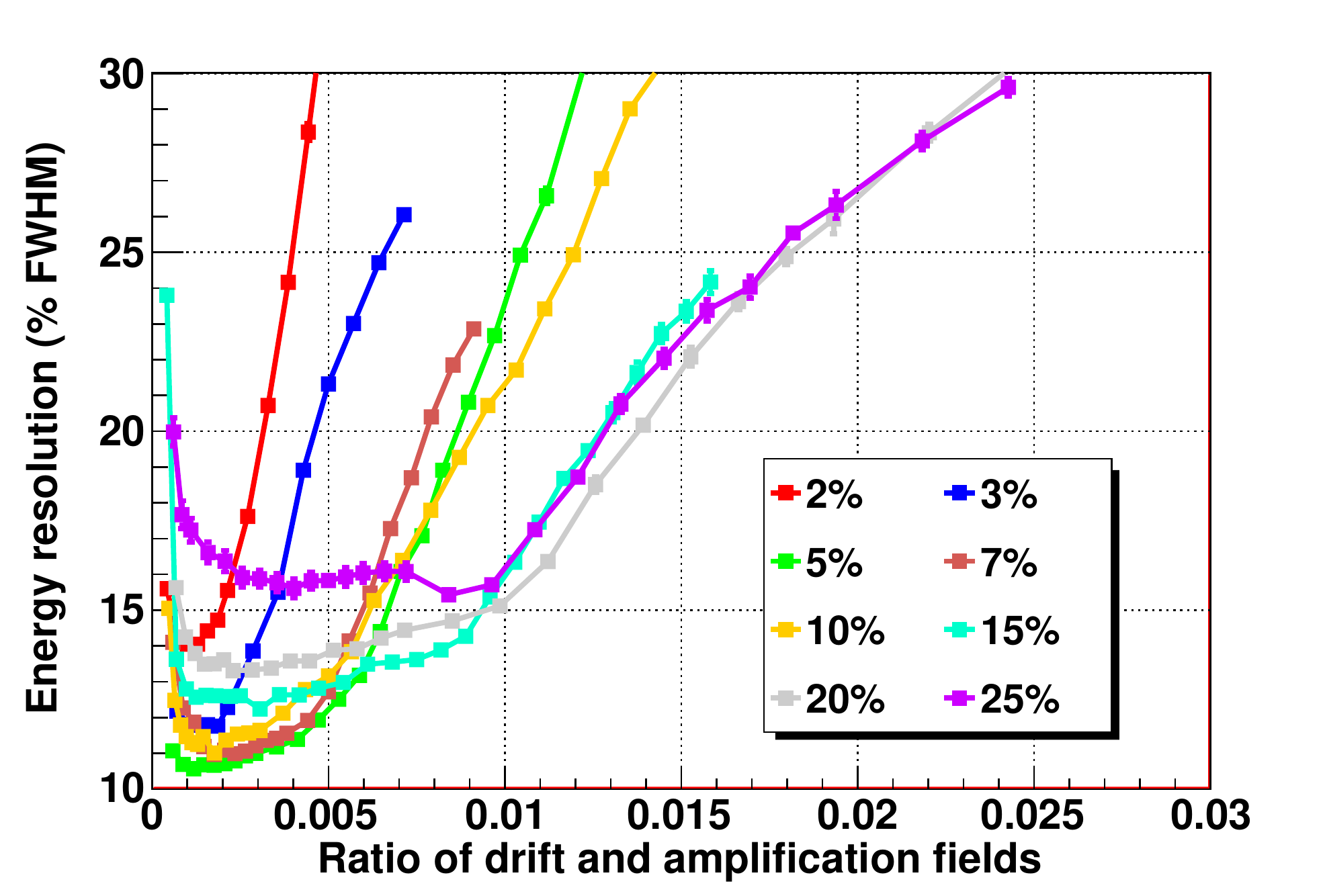}
\caption{\it Dependence of the electron transmission (left) and the energy resolution (right) with the ratio of drift and amplification fields for a microbulk detector of 50~$\mu$m-thickness gap in neon-isobutane mixtures. The gain has been normalized to the maximum of each series. The percentage of each series corresponds to the isobutane concentration.}
\label{fig:TransEResNeIso}
\end{figure}

\medskip
The electron transmission curves of both detectors show the same features observed in argon-based mixtures. For the case of the 50~$\mu$m-thickness gap, the plateau of maximum mesh transparency widens as the quencher concentration is increased, as shown in figure \ref{fig:TransEResNeIso} (left) for isobutane, and the energy resolution is correlated to this variable (same figure, right). The plateau is absent for the other detector but there is a narrow range of fields for which the energy resolution reaches its best values. Comparing their performances to argon-based mixtures, we have observed narrower plateaus of maximum transmission for the gap of 50~$\mu$m, which can be explained by the higher diffusion coefficients in neon \cite{Magboltz}.

\begin{figure}[htb!]
\centering
\includegraphics[width=75mm]{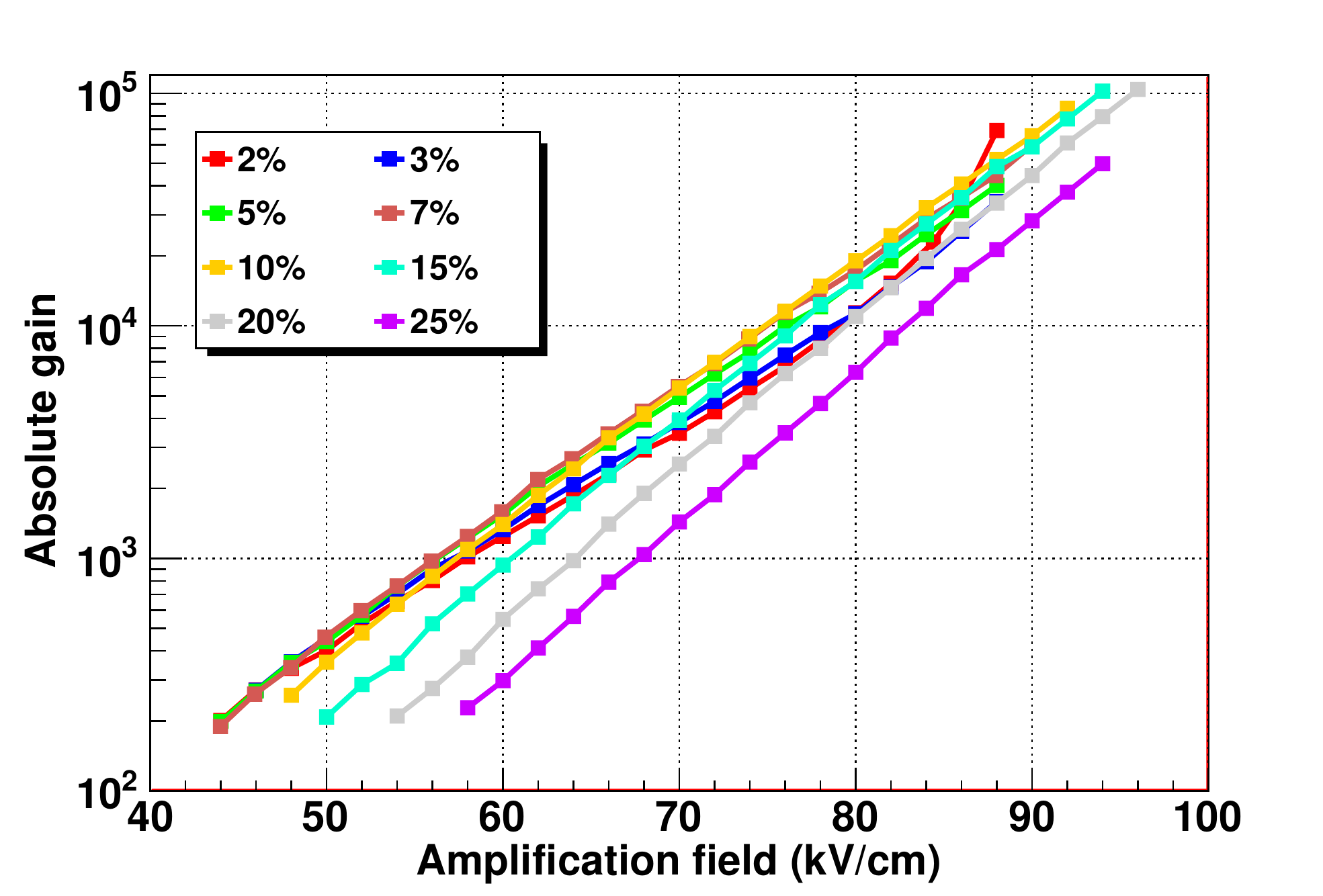}
\includegraphics[width=75mm]{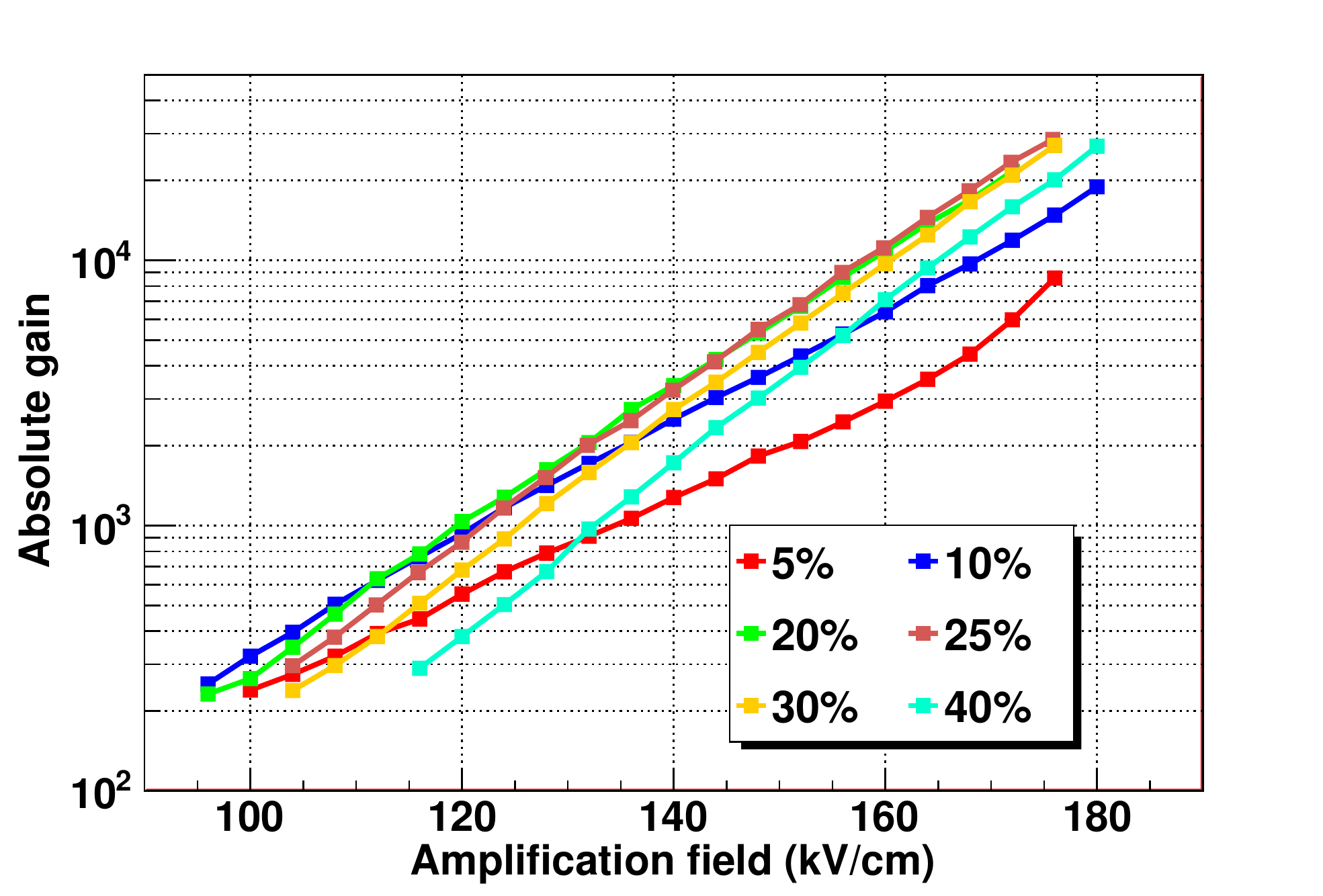}
\caption{\it Dependence of the absolute gain with the amplification field for two microbulk detectors with gaps of 50 (left) and 25~$\mu$m (right) in neon-isobutane mixtures. The maximum gain of each curve was obtained just before the spark limit. The percentage of each series corresponds to the isobutane concentration.}
\label{fig:GainNeIso}
\end{figure}

\medskip
The dependence of the absolute gain with the amplification field is shown in figure \ref{fig:GainNeIso}. Gains higher than $5 \times 10^4$ and $2 \times 10^4$ are respectively reached for isobutane concentrations greater than 7 and 10\% and for a gap of 50 and 25~$\mu$m. Moreover, the required field for a fixed gain does not increase with the quencher concentrations below 10\%, as in argon-based mixtures. We will try to interpret this difference in terms of gas properties in the next section. A second fact observed is that higher fields are needed in these mixtures to get the same gain. As an example, a field of 72 kV/cm must be applied in Ar+10\% iC$_4$H$_{10}$ to reach a gain of 10$^4$ for the detector of 50~$\mu$m-thickness gap. In a neon based mixture, this value is 75 kV/cm. This difference is greater for the other detector: respectively 140 kV/cm and 167 kV/cm for the same mixtures.

\medskip
An improvement in the energy resolution has been observed for neon-based mixtures and the 50~$\mu$m-thickness-gap detector, going from 11.7\% FWHM for Ar+5\%iC$_4$H$_{10}$ down to 10.5\% FWHM for Ne+7\%iC$_4$H$_{10}$. This fact cannot be explained by the fluctuations of the primary ionization but by the contribution of the detector avalanches. As described in \cite{Blum:1994wb}, the energy resolution of the detector can be expressed as
\begin{equation}
R(\mbox{\% FWHM}) \ = \ 2.35 \ \sqrt{\frac{W}{E_0} \ (F + b)}
\end{equation}

where $E_0$ is the event energy, $F$ is the gas Fano Factor, $W$ is the mean energy necessary for creating an electron-ion pair in the gas and $b$ is the detector contribution. In neon-based mixtures, less primary electrons are generated due to its higher mean electron-ion pair energy (respectively 36.4 and 26.3~eV for neon and argon \cite{Blum:1994wb}), which should worsen the energy resolution as the Fano factor is similar to both gases (0.17 \cite{Biagi:2011}). The detector contribution must be thus little for neon-based mixtures. In fact, as noted in \cite{Schindler:2010hs}, less avalanche fluctuations are presented in light gases like helium or neon due to their higher ionization yield.

\begin{figure}[htb!]
\centering
\includegraphics[width=75mm]{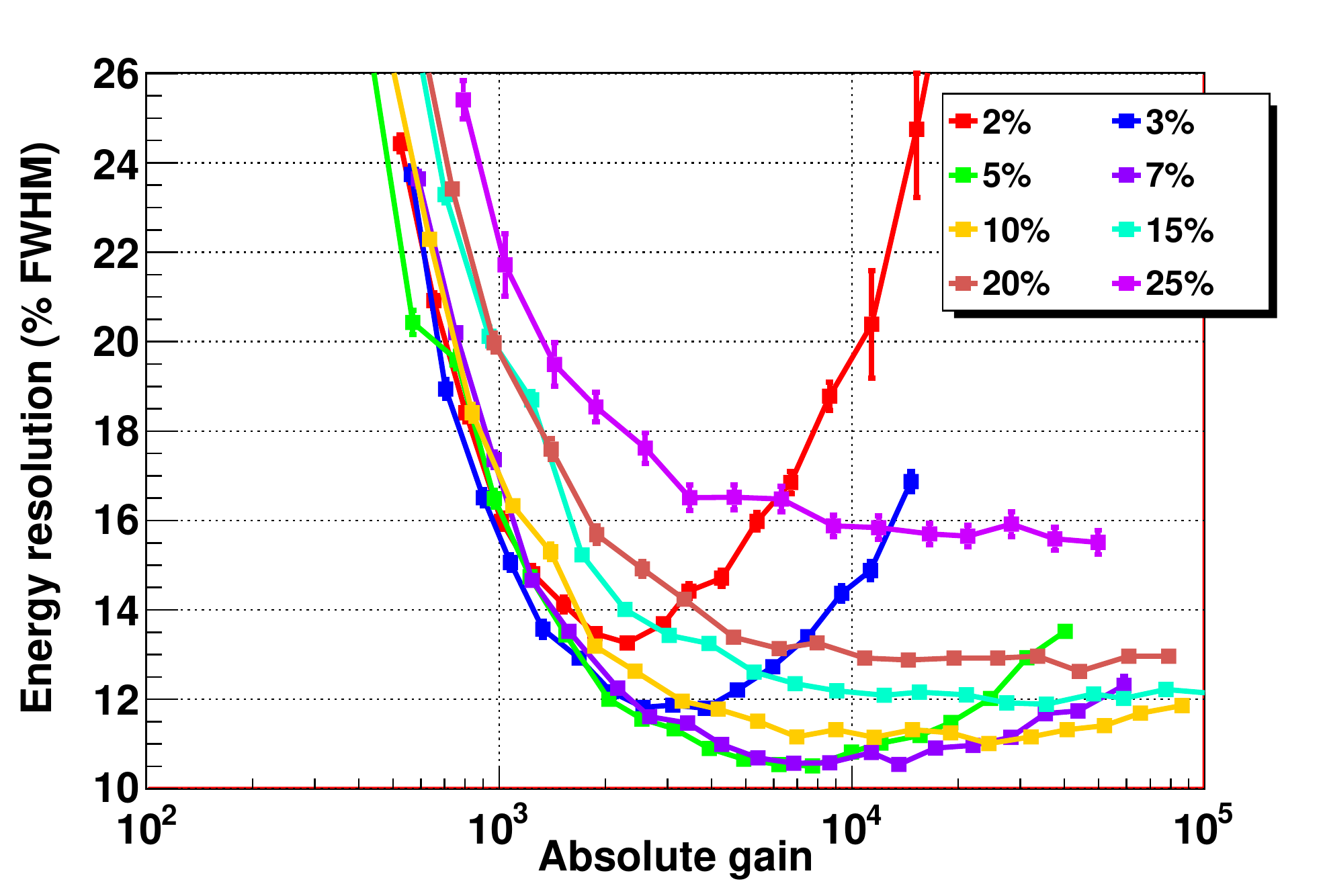}
\includegraphics[width=75mm]{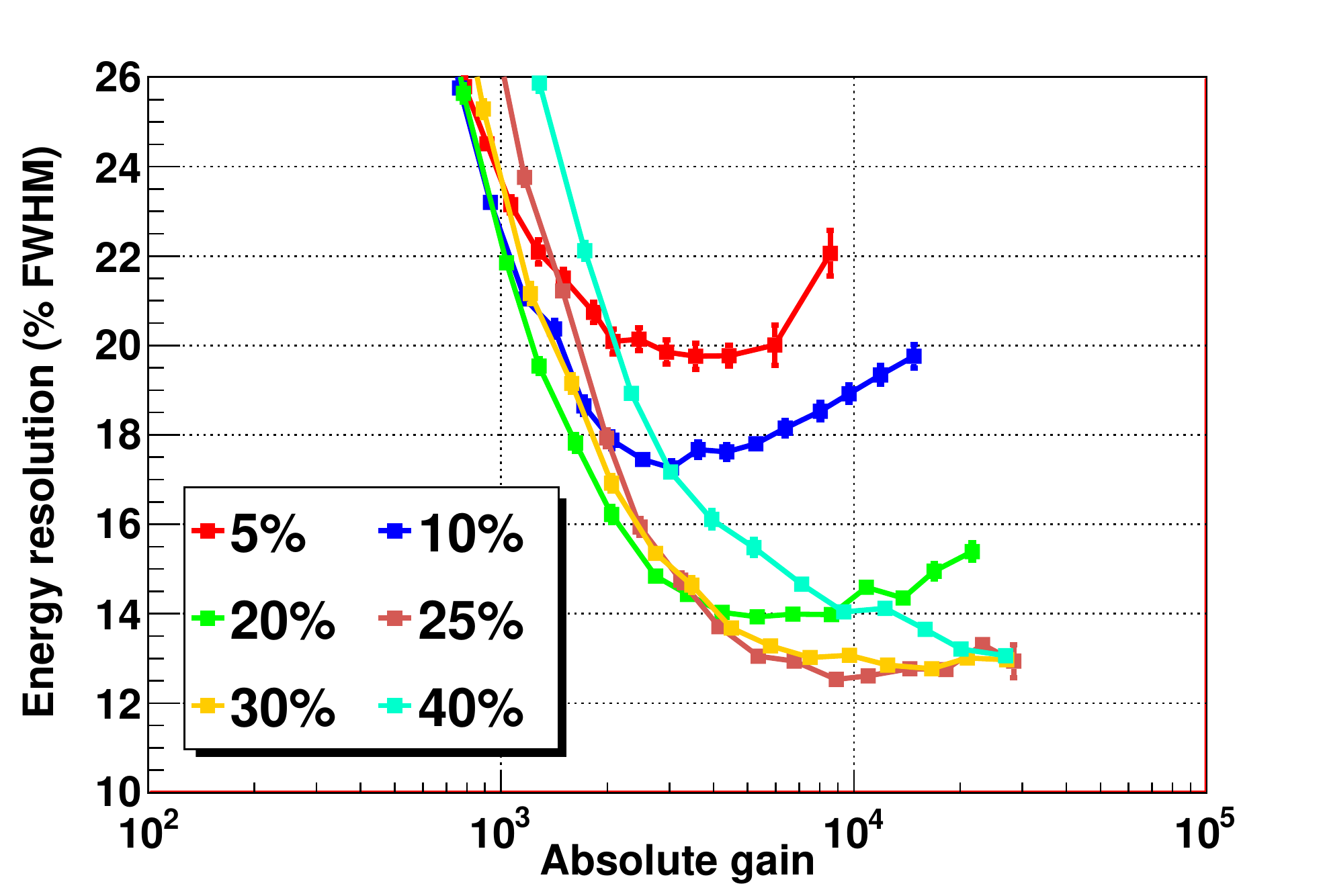}
\caption{\it Dependence of the energy resolution with the absolute gain for two detectors of 50 (left) and 25~$\mu$m-thickness-gap (right) in neon-isobutane mixtures. The maximum gain of each curve was obtained just before the spark limit. The percentage of each series corresponds to the isobutane concentration.}
\label{fig:EResGainNeIso}
\end{figure}

\medskip
The energy resolution is also better at high gains, as shown in figure \ref{fig:EResGainNeIso} (left). At a gain of $5 \times 10^4$, the energy resolution is 11.4\% FWHM in Ne+10\%iC$_4$H$_{10}$, better than the 15\%~FWHM reached in Ar+11\% cyclohexane. Similar values are also obtained with ethane or cyclohexane, as shown in table \ref{tab:eresneon}, which may point out that it is mainly determined by the base gas.

\medskip
In contrast, the energy resolution is worse in neon-based mixtures for the 25~$\mu$m-thickness-gap detector, passing from 11.6\% FWHM in Ar+10\%iC$_4$H$_10$ to 12.6\% FWHM in Ne+25\%iC$_4$H$_{10}$. Worse values have also been obtained with the other quenchers, as shown in table \ref{tab:eresneon}. Larger amplification gaps are in principle more adequate for light gases as they minimize the avalanche fluctuations produced by gap uniformaties \cite{Giomataris:1998yg}. Another possible explanation is a small energy transfer between neon and the quencher due to the gap size \cite{Schindler:2011hs}, which may limit the Penning effect.

\begin{table}[htb!]
\begin{center}
$$
\begin{array}{cc|cc|cc}
\multicolumn{1}{c}{}&\multicolumn{1}{c}{}&\multicolumn{2}{c}{\mbox{50~$\mu$m gap}}&\multicolumn{2}{c}{\mbox{25~$\mu$m gap}}\\
{\mbox{Type of}}&{\mbox{Ion. thres}}&{\mbox{Resolution}}&{\mbox{Quencher}}&{\mbox{Resolution}}&{\mbox{Quencher}}\\
{\mbox{quencher}}&{\mbox{eV}}&{\mbox{\% FWHM}}&{\%}&{\mbox{\% FWHM}}&{\%}\\
\hline
{\mbox{Ethane}}&{11.7}&{10.7}&{5-15}&{14.8}&{25}\\
{\mbox{Isobutane}}&{10.8}&{10.5}&{5-7}&{12.6}&{25-30}\\
{\mbox{Cyclohexane}}&{9.9}&{10.6}&{8-10}&{16.6}&{10}\\
\end{array}
$$
\end{center}
\caption{\it Best values of the energy resolution in mixtures of neon with ethane, isobutane and cyclohexane for two microbulk detectors with gaps of 50 and 25~$\mu$m. The range of quencher concentration for which these values are obtained is also shown. The systematic error of all values is 0.2\% FWHM.}
\label{tab:eresneon}
\end{table}

\begin{figure}[htb!]
\centering
\includegraphics[width=75mm]{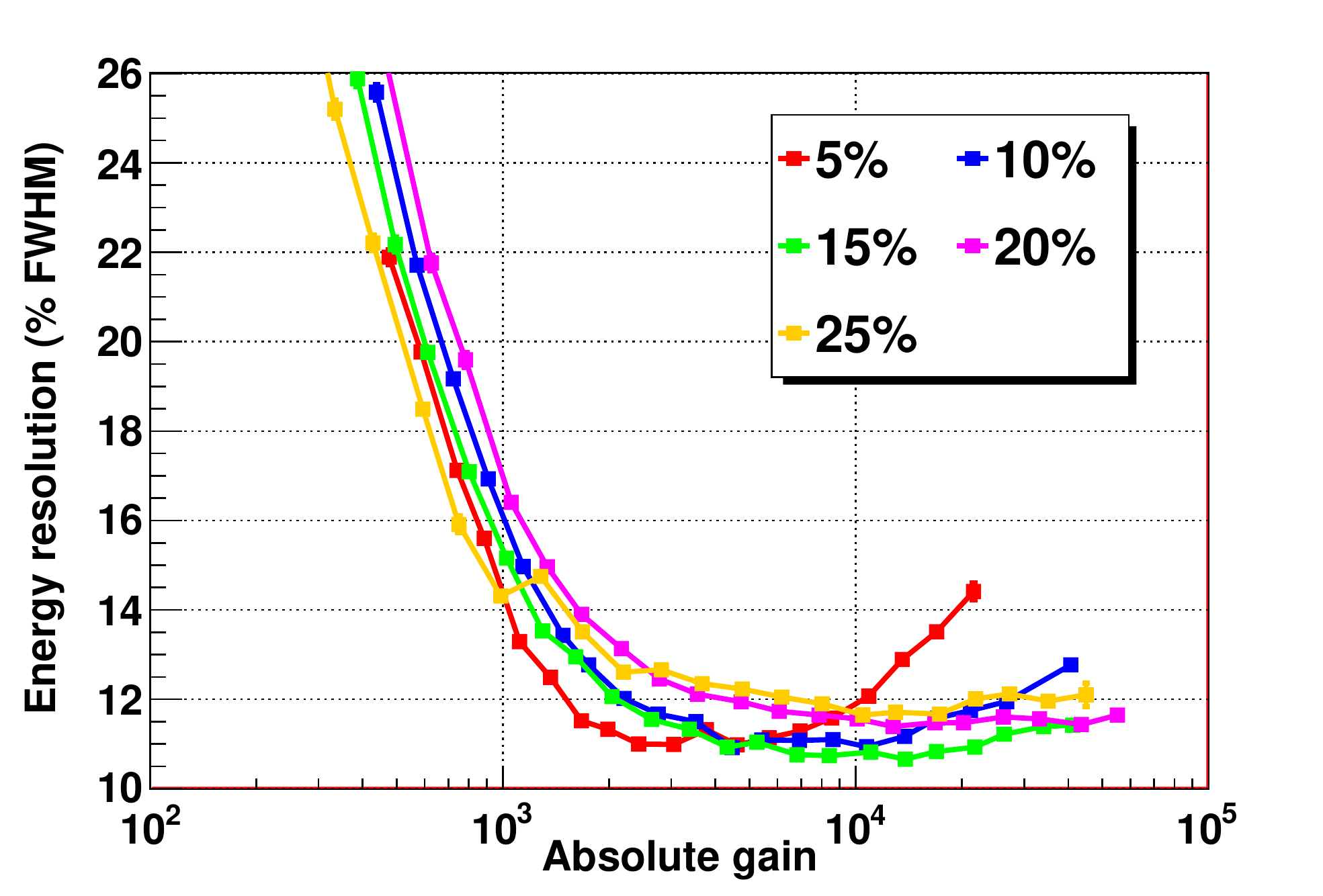}
\includegraphics[width=75mm]{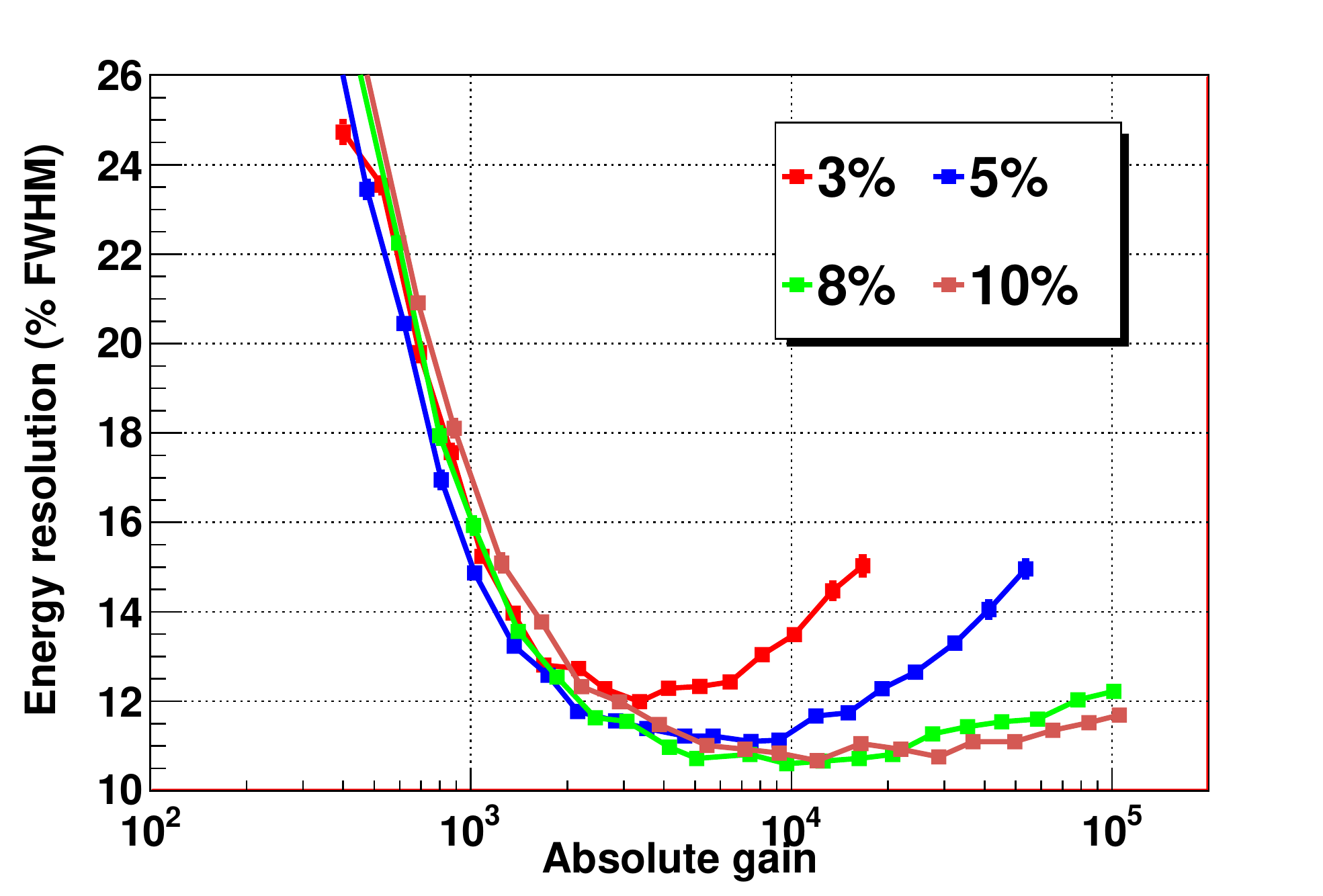}
\caption{\it Dependence of the energy resolution with the absolute gain for a detector of 50~$\mu$m-thickness-gap in neon-ethane (left) and -cyclohexane mixtures (right). The maximum gain of each curve was obtained just before the spark limit. The percentage of each series corresponds to the quencher concentration.}
\label{fig:EResGainNeM50}
\end{figure}

\medskip
We finally present in figure \ref{fig:EResGainNeM50} the dependence of the energy resolution with the gain for the microbulk of 50~$\mu$m gap in mixtures of neon with ethane and cyclohexane. The maximum gain before the spark limit is respectively $5 \times 10^4$ and $10^5$. In contrast to argon-based mixtures, there is no degradation of the energy resolution at high gains.

\section{Micromegas and the Rose-Korff gain model}
\begin{figure}[htb!]
\centering
\includegraphics[width=70mm]{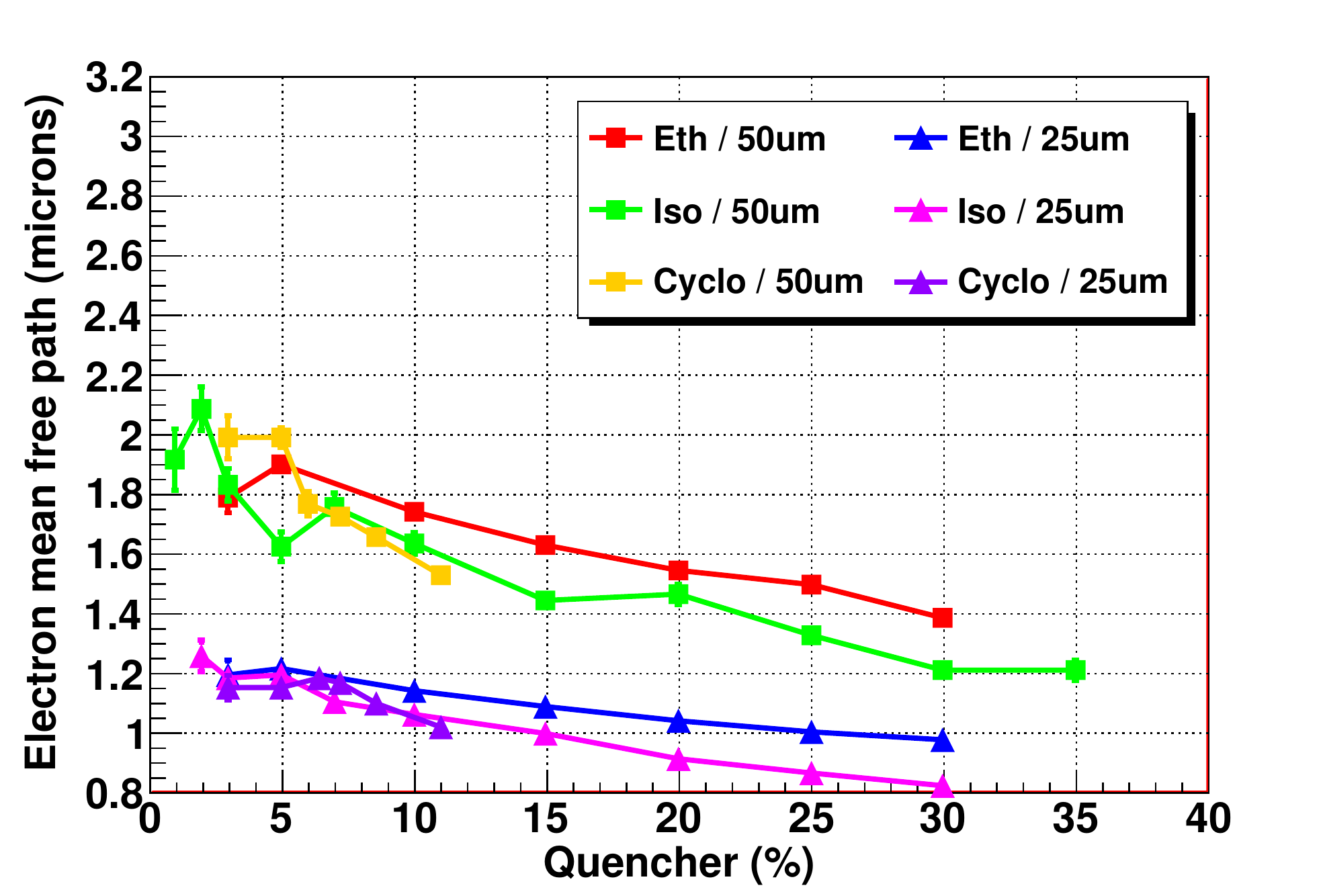}
\includegraphics[width=70mm]{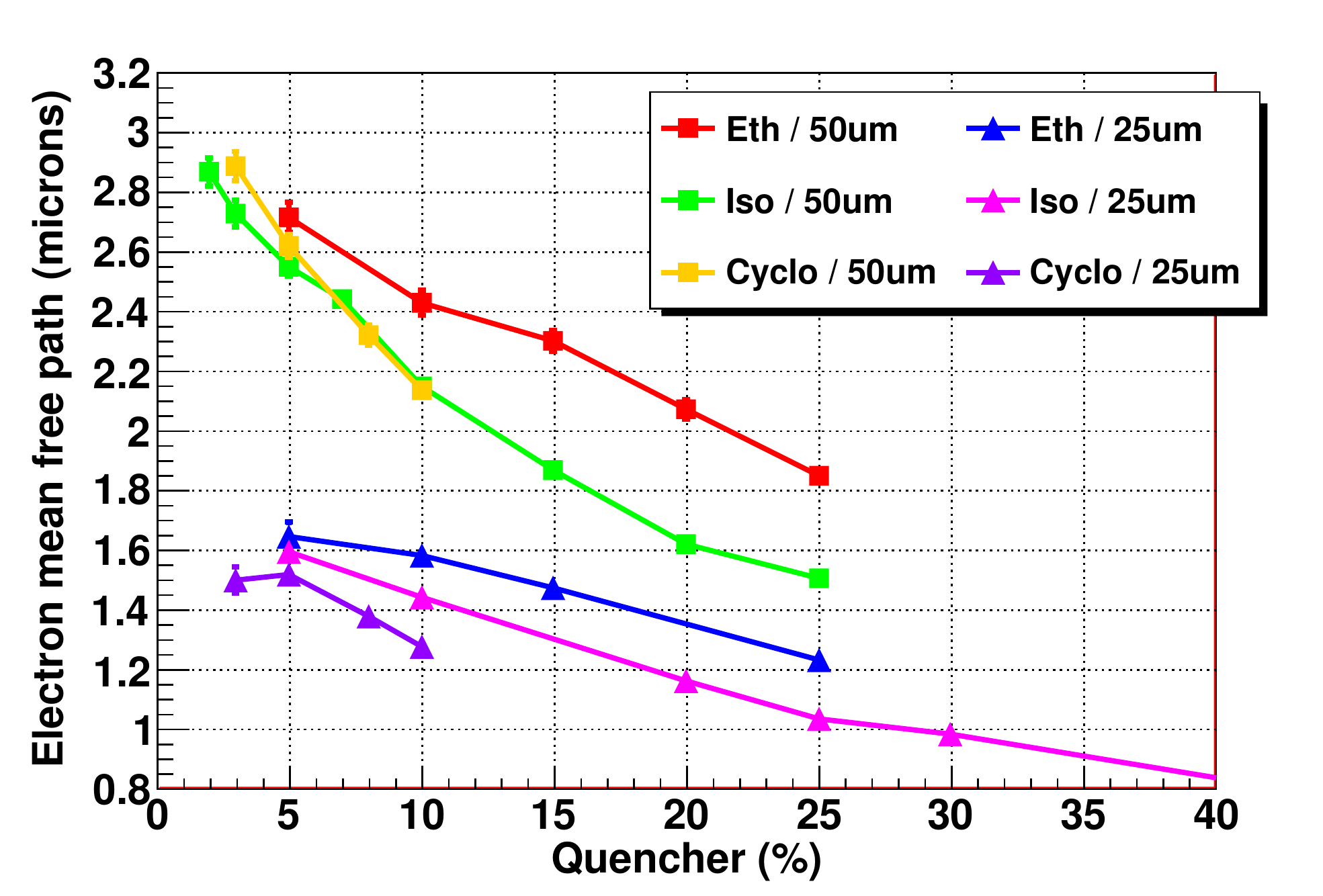}
\caption{\it Dependence of the electron mean free path with the quencher concentration for the argon- (left) and neon-based (right) mixtures studied. Values have a relative systematic error of 10\%.}
\label{fig:FitGainMFP}
\end{figure}

\begin{figure}[htb!]
\centering
\includegraphics[width=70mm]{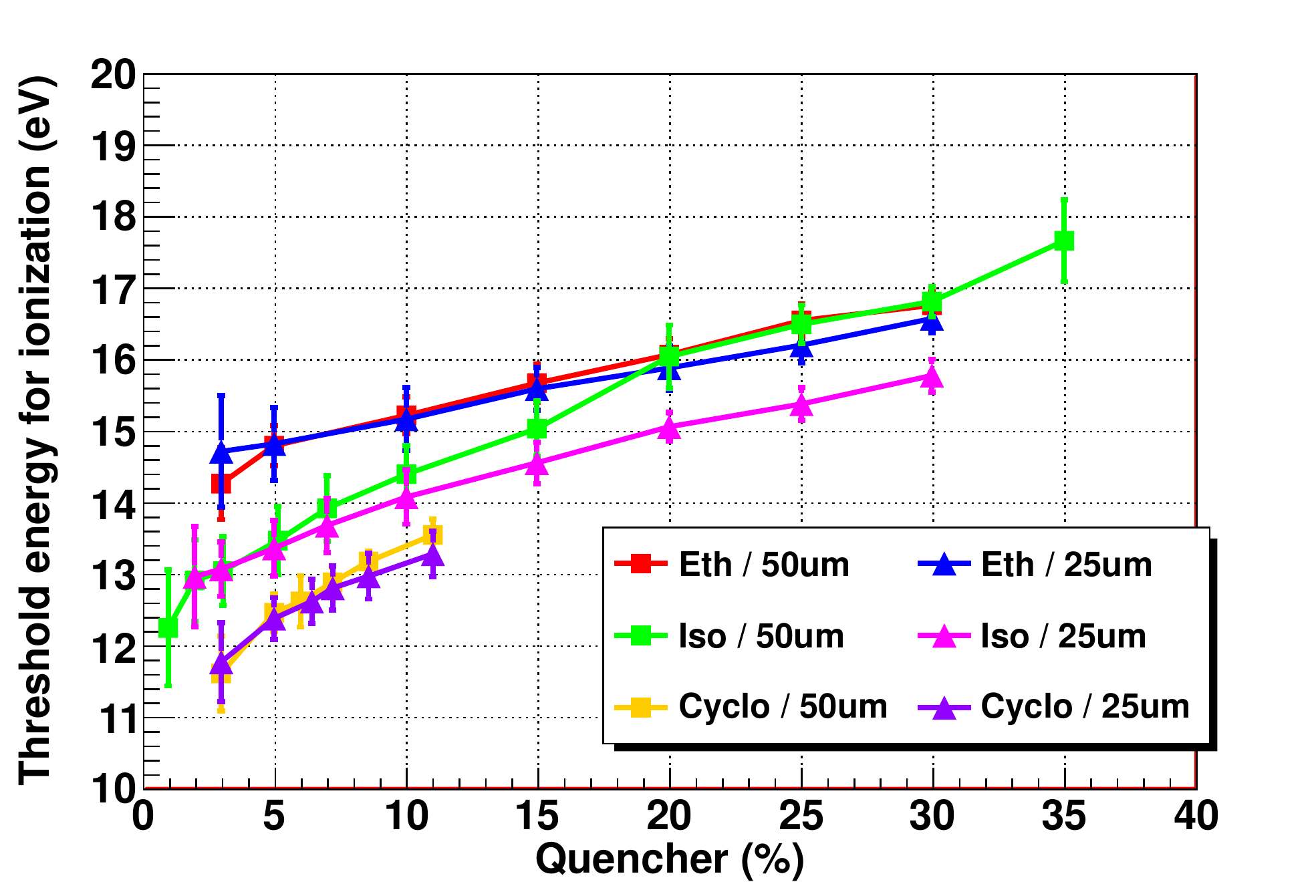}
\includegraphics[width=70mm]{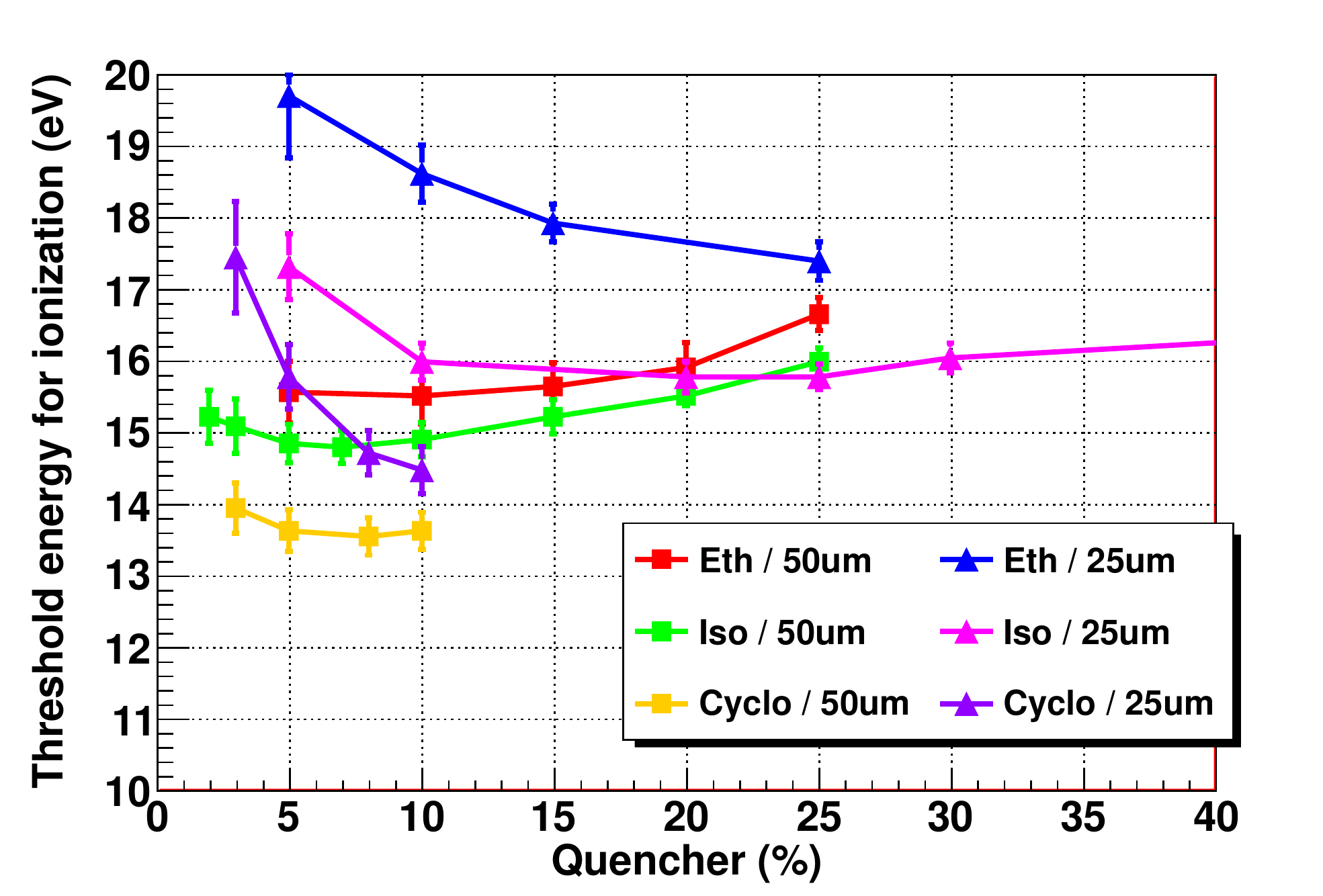}
\caption{\it Dependence of the energy threshold for ionization with the quencher concentration for the argon- (left) and neon-based (right) mixtures studied. Values have a systematic error of 10\%.}
\label{fig:FitGainIe}
\end{figure}

%%% Introduction and presentation about the Rose-Korff gain model.
The Rose-Korff model is a parametrization of the first Townsend coefficient which neglects the possible secondary ionizations in the avalanche \cite{Rose:1941mr}. If we modelize Micromegas as a parallel-plate detector and we suppose that the amplification field is uniform, the gain can be expressed by the formula
\begin{equation}
ln(G) = \frac{d}{\lambda} \exp\left(\frac{-I_e}{\lambda \ E_{amp}}\right)
\label{eq:rose}
\end{equation}

where $G$ is the gain, $E_{amp}$ is the amplification field, $d$ is the gap distance, $\lambda$ is the electron mean free path and $I_e$ is the threshold energy for ionization. Therefore, this simple model can give information on gas properties. This study was firstly made in \cite{Attie:2000}. The gain curves generated in argon- and neon-based mixtures have been fitted to the equation \ref{eq:rose} and the resulting dependencies of the electron mean free path and the threshold energy for ionization are respectively shown in figures \ref{fig:FitGainMFP} and \ref{fig:FitGainIe}.

\medskip
As shown in figure \ref{fig:FitGainMFP}, the electron mean free path decreases as the quencher concentration is increased and, for the same base gas and detector, there is almost no difference for low quencher quantities. At high concentration, mixtures with ethane show higher values than with isobutane. Apart from that, neon-based mixtures show higher values than in argon-based ones. For instance, the electron mean free path is respectively 1.6 and 2.5~$\mu$m in Ar+5\%iC$_4$H$_{10}$ and Ne+5\%iC$_4$H$_{10}$ for the 50~$\mu$m-thickness-gap detector. These three facts can be explained by theory \cite{Blum:1994wb}, even if the exact values do not match. The electron mean free path is in general higher for light gases, for noble gases in comparison to hydrocarbons and for ethane compared to isobutane. Finally, we have also observed lower values for the gap size of 25~$\mu$m, which cannot be explained by gas properties.

\begin{figure}[htb!]
\centering
\includegraphics[width=70mm]{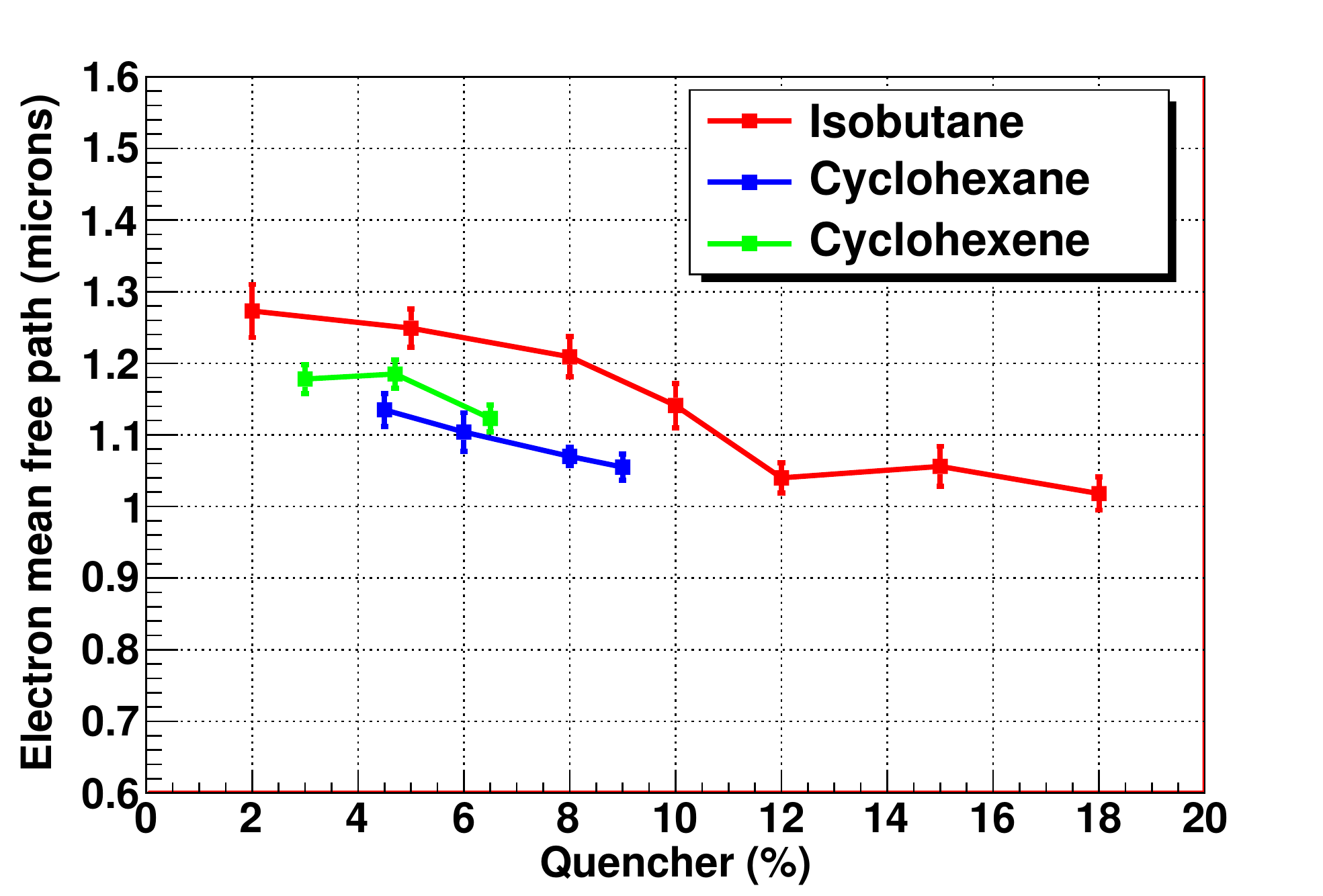}
\includegraphics[width=70mm]{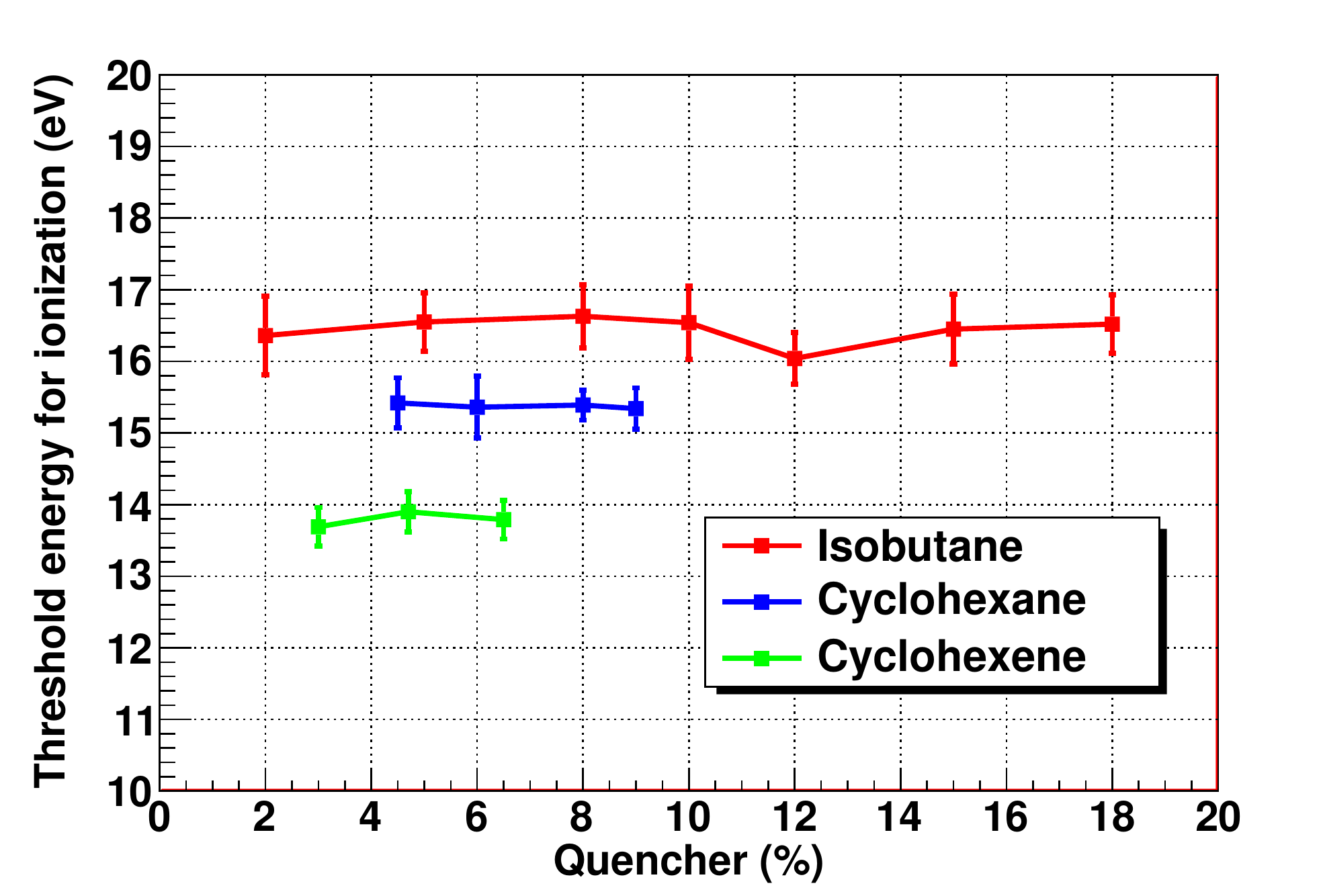}
\caption{\it Dependence of the electron mean free path (left) and the energy threshold for ionization (right) with the quencher concentration for different xenon-based mixtures generated by the first micromegas detectors with a gap of 50~$\mu$m \cite{Laval:1997lv}. Values have a relative systematic error of 20\%.}
\label{fig:FitGainXenon}
\end{figure}

\medskip
To confirm experimentally that the electron mean free path is shorter for heavier gases, we have revisited the gain curves generated for xenon-based mixtures in \cite{Laval:1997lv}. The electron mean free path (figure \ref{fig:FitGainXenon}, left) is clearly lower than in other gases, as xenon is a heavier gas. Small differences are due to higher systematic errors in the gain measurements.

\medskip
The energy threshold for ionization has a different behaviour in argon- than in neon-based mixtures, as shown in figure \ref{fig:FitGainIe}. For the first type, this parameter increases with the quencher quantity and is independent of the gap size. For example, for a quencher percentage of 10\%, the energy threshold is respectively 15.2, 14.3 and 13.5 eV for ethane, isobutane and cyclohexane. In contrast, this parameter shows a minimum in neon-based mixtures, which depends on both the quencher and the detector. For all mixtures, values are higher for a gap size of 25~$\mu$m.

\medskip
These differences can be explained by the Penning effect. The transfer of energy from non-metastable levels of argon to the quencher molecule is complete for both detectors in argon-based mixtures and the energy threshold is only defined by the quencher. For neon, as the electron mean free path is longer, the transfer is limited by the gap size. This limitation disappears at high quencher quantities and the energy threshold for ionization increases with quencher concentration, as in argon-based mixtures.

\medskip
Finally, the values obtained in neon are in general higher than in argon-based mixtures. For example, this parameter is respectively 15.6 and 14.8 eV in Ne+5\% ethane and Ar+5\% ethane for the 50~$\mu$m-thickness-gap detector. This fact is due to the higher values of this parameter for pure neon (21.6 eV) than for pure argon (15.8 eV) \cite{Blum:1994wb}. Note that the values given in literature for the base gases and for the quenchers (11.7 and 10.8 eV for ethane and isobutane) do not match with the ones obtained. As discussed in \cite{Attie:2000, Guo:2010jjg}, this fact can be explained by the Penning effect \cite{Bronic:1996ikb}, which is quite important in mixtures with big differences between the ionization threshold of the quencher and the excitation levels of the base gas.

\medskip
A good question is if the values obtained for the electron mean free path and the energy threshold for ionization could explain the energy resolution of a detector in a gas. We can start our discussion using \cite{Giomataris:1998yg}, where it was concluded that the fluctuations due to defects of the mesh and the gap flatness could be compensated chosing the idoneous gap for each gas type. This optimum was obtained by differentiating the equation \ref{eq:rose}, which results in
\begin{equation}
\frac{\delta G}{G} = G \ \left(1 - \frac{d \ I_e}{\lambda \ V_{amp}}\right) \frac{\delta d}{d} = 0 \Rightarrow E^{opt}_{amp} = \frac{I_e}{\lambda}
\end{equation}

\medskip
We can interpret this minimum in terms of an optimum amplification field for a fixed gap. For each gas, this field should be as low as possible to fastly reach an optimum detector performance and to avoid the degradation of the energy resolution by other effects. Therefore, the electron mean free path should be as high as possible and the values of the energy threshold for ionization should be little.

\medskip
As shown in figures \ref{fig:FitGainMFP} and \ref{fig:FitGainXenon} (right), the first parameter is mainly determined by the base gas, being higher for neon than for argon or xenon. The quencher concentration decreases its value, which in general degrades the energy resolution. There is also a dependence on the gap size, as lower values are obtained for a gap of 25~$\mu$m. As we have shown, the gap size is only important in neon-based mixtures. In the case of the energy threshold for ionization, we have observed lower values for cyclohexane than for isobutane and ethane but there is no clear improvement in the energy resolution. This variable could have thus little effect if the optimum amplification field is in the working range.

\section{Conclusions and perspectives}
Microbulk is a Micromegas technology that has been used in nuclear (nTOF) and astroparticles experiments (CAST, NEXT). It offers uniform and flexible structures with an excellent energy and time resolution, low background levels and low mass. Two detectors, with a gap distance of respectively 50 and 25~$\mu$m, have been characterized in argon-based mixtures with ethane, isobutane and cyclohexane, using a $^{55}$Fe source. The maximum gain before the spark limit is respectively $2 \times 10^4$ and $10^4$ and the best energy resolutions obtained in argon-isobutane mixtures are 11.7\% and 12.1\%~FWHM at 5.9 keV. The plateau of maximum electron transmission depends on the quantity of quencher for the 50~$\mu$m-thick-gap detector and is almost inexistent for the gap of 25~$\mu$m. The same conclusions have been obtained with the other quenchers but the gain is higher in argon-cyclohexane mixtures and lower with ethane.

\medskip
Microbulk detectors have been also tested in neon-based mixtures with the same quenchers to reduce the energy threshold and to make Micromegas sensitive to sub-keV energies. Possible applications are synchroton radiation and Dark Matter searches. The maximum gain reached before the spark limit in argon-isobutane mixtures is respectively $5 \times 10^4$ and $10^5$ for a gap size of 25 and 50~$\mu$m and the best energy resolutions are 10.5\% and 12.6\%~FWHM at 5.9 keV. Similar values for the energy resolution have been obtained with the other quenchers except for the detector of 25~$\mu$m-thickness gap and neon-based mixtures, whose values are worse.

\medskip
The last developments of microbulk technology are focused in creating detectors with small gaps (12.5 and 25~$\mu$m) to be used with gases at high pressure. Different hole pitchs and diameters are being studied. The use of other quenchers with even lower ionization threshold like cyclohexene are also planned, in order to confirm the divergences observed in argon-cyclohexane between micromegas detectors and proportional counters.
%%%%%%%%%%%%%%%%%%%%%%%%%%%%%%%%%%%%%%%%%%%%%%%%%%%%%%%%%%%%%%%%%%%%%%%%%
\section*{Acknowledgements}
We are thankfull to our colleagues of the University of Zaragoza and IRFU/CEA-Saclay, as well as to the RD51 collaboration for helpful discussions and encouragements. F. I. acknowledge supports from the Eurotalents program.

%%%%%%%%%%%%%%%%%%%%%%%%%%%%%%%%%%%%%%%%%%%%%%%%%%%%%%%%%%%%%%%%%%%%%%%%%


\begin{thebibliography}{00}
%% Micromegas technology.
\bibitem{Giomataris:1995fq} 
  Y.~Giomataris {\it et al.},
  MICROMEGAS: A high - granularity position - sensitive gaseous detector for high particle - flux environments,
  {\it Nucl.\ Instrum.\ Meth.\  A} {\bf 376} (1996) 29.
%% Bulk technology.
\bibitem{Giomataris:2006yg}
  I. Giomataris {\it et al.},
  Micromegas in a bulk,
  {\it Nucl. Instrum. Meth. A} {\bf 560} (2006) 405.
%% Microbulk technology
\bibitem{Adriamonje:2010sa}
  S.~Adriamonje {\it et al.},
  Development and performance of Microbulk Micromegas detectors,
  {\it JINST} {\bf 5} (2010) P02001.
%%% Another reference to Microbulk technology..
\bibitem{Iguaz:2011fi}
  F.J.~Iguaz {\it et al.},
  New developments of Micromegas Microbulk technologies,
  Proceeding of the TIPP 2011 conference, {\it Preprint:} arXiv:1110.2641.
%% Radiopurity of Microbulk detectors.
\bibitem{Cebrian:2011sc}
  S.~Cebrian {\it et al.},
  Radiopurity of Micromegas readout planes,
  {\it Astropart. Phys.} {\bf 34} (2011) 354.
%% Neutron measurements with n_TOF.
\bibitem{Colonna:2011nc}
  N.~Colonna {\it et al.},
  Neutron reasurements for advanced nuclear systems: The n\_TOF project at CERN,
  Accepted in {\it Nucl. Instrum. Meth. B}.
%%% Micromegas detectors in CAST.
\bibitem{SAune:2009sa}
  S.~Aune {\it et al.},
  An ultra-low background detector for axion searches,
  {\it J. Phys. Conf. Ser.} {\bf 179} (2009) 012015.
%%%% Double beta decay studies.
%% Micromegas in pure xenon.
\bibitem{Cebrian:2010sc2}
  S.~Cebrian {\it et al.},
  Micromegas readouts for double beta decay searches,
  {\it JCAP} {\b 10} (2010) 1010.
%% Talk by Theopisti in TPC2010.
\bibitem{Dafni:2010td}
  T.~Dafni {\it et al.},
  Micromegas planes for the neutrinoless double beta decay search with NEXT,
  {\it J. Phys. Conf. Ser.} {\bf 309} (2011) 012009.
%% Optimum for argon-based mixtures.
\bibitem{Agrawal:1988pca}
  P.C.~Agrawal and B.D.~Ramsey,
  Use of propane as a quench gas in argon-filled proportional counters and comparison with other quench gases,
  {\it Nucl. Instrum. Meth. A} {\bf 273} (1988) 331.
%%% An optimum gap for each type of gas and pressure.
\bibitem{Giomataris:1998yg}
  Y. Giomataris,
  Development and prospects of the new gaseous detector Micromegas,
  {\it Nucl. Instrum. Meth. A} {\bf 419} (1998) 239.
%%% H. Schindler opinion.
\bibitem{Schindler:2011hs}
  H.~Shindler,
  private communication.
%% Microbulks en xenon puro.
\bibitem{Dafni:2009df}
  Th. Dafni {\it et al.},
  Energy resolution of alpha particles in a Micromegas detector at high pressure,
  {\it Nucl. Instrum. Meth. A} {\bf 608} (2009) 259.
%% Microbulks en argon puro.
\bibitem{Iguaz:2010fj}
  F.J.~Iguaz {\it et al.},
  New results of microbulk detectors,
  talk in the 5th RD51 Collaboration Metting.
%% Tesis de Max Chefdeville
\bibitem{Chefdeville:2009mc}
  M. Chefdeville,
  Development of Micromegas-like gaseous detectors using a pixel readout chip as collecting anode,
  PhD Thesis, University of Amsterdam, 2009.
%% Desviacion del modelo Rose-Korff de ganancia.
\bibitem{Bronic:1998ikb}
  I.K.~Bronic and B.~Grosswendt,
  Gas amplification and ionization coefficients in isobutane and argon-isobutane mixtures at low gas pressures,
  {\it Nucl. Instrum. Meth. B} {\bf 142} (1998) 219.
%% Fluctuaciones en la ganancia.
\bibitem{Schindler:2010hs}
  H.~Schindler {\it et al.},
  Calculation of gas gain fluctuations in uniform fields,
  {\it Nucl. Instrum. Meth. A} {\bf 624} (2010) 78.
%% Rose and Korff gain model.
\bibitem{Rose:1941mr}
  M.E.~Rose and S.A.~Korff,
  An Investigation of the Properties of Proportional Counters. I,
  {\it Phys. Rev.} {\bf 59} (1941) 850.
%% Optimum for xenon-based mixtures.
\bibitem{Agrawal:1989pca}
  B.D.~Ramsey and P.C.~Agrawal,
  Xenon-based peenning mixtures for proportional counters,
  {\it Nucl. Instrum. Meth. A} {\bf 278} (1989) 576.
%% Reference to Magboltz
\bibitem{Magboltz}
  Magboltz - transport of electrons in gas mixtures,
  http://magboltz.web.cern.ch/magboltz/.
%% Penning effect by non-metastable states.
\bibitem{Sahin:2010os}
  \"O.~Sahin {\it et al.},
  Penning transfer in argon-based gas mixtures,
  {\it JINST} {\bf 5} (2010) P05002.
%% Reference to a model with Penning mixtures.
\bibitem{Bronic:1996ikb}
  I.K.~Bronic and B.~Grosswendt,
  Ionization yield formation argon-isobutane mixtures as measured by proportional-counter method,
  {\it Nucl. Instrum. Meth. B} {\bf 117} (1996) 5.
%% Blum and Rolandi reference.
\bibitem{Blum:1994wb}
  W. Blum and L. Rolandi,
  Particle Detection with Drift Chambers,
  Springer-Verlag, 1993.
%% Reference to Fano factor
\bibitem{Biagi:2011}
  S.~Biagi,
  MIP program,
  http://consult.cern.ch/writeup/magboltz/.
%% Reference to David Attie presentation.
\bibitem{Attie:2000}
  D.~Atti\'e {\it et al.},
  Gain measurements in various gas mixtures and optimization of the stability,
  talk at the TPC Jamboree, Aachen, 2007.
%% Reference to the Chinese article on Micromegas.
\bibitem{Guo:2010jjg}
  J.J.~Guo {\it et al.},
  3D simulation of micromegas detector performance,
  {\it Chinese Physics C} {\bf 34} (2010) 482.
%% Laval report.
\bibitem{Laval:1997lv}
  L.~Laval,
  Etude de melanges gazeux pour l'optimisation du d\'etecteur Micromegas,
  stage report at the CEA/DSM, 1997.
\end{thebibliography}
\end{document}